\documentclass[sigconf]{acmart}

\usepackage{amssymb}
\usepackage{pifont}
\usepackage{makecell}
\usepackage{multirow}
\usepackage{graphicx}
\usepackage{subcaption}
\usepackage{xcolor}
\usepackage{enumitem} 

\AtBeginDocument{%
  }

\setcopyright{acmlicensed}
\copyrightyear{2018}
\acmYear{2018}
\acmDOI{XXXXXXX.XXXXXXX}
\acmConference[Conference acronym 'XX]{xxx}{June 03--05,
  2018}{Woodstock, NY}
\acmISBN{978-1-4503-XXXX-X/2018/06}

\begin{document}

\title{Similar Users-Augmented Interest Network}

\author{Xiaolong Chen}
\email{chenxiaolong@mail.ustc.edu.cn}
\affiliation{%
  \institution{University of Science and Technology of China}
  \country{China}
}
\authornote{Equal contribution.}

\author{Haoyi Zhao}
\email{asteriazhao@mail.ustc.edu.cn}
\affiliation{%
  \institution{University of Science and Technology of China}
  \country{China}
}
\authornotemark[1]

\author{Xu Huang}
\email{xuhuangcs@mail.ustc.edu.cn}
\affiliation{%
  \institution{University of Science and Technology of China}
  \country{China}
}

\author{Defu Lian}
\email{liandefu@ustc.edu.cn}
\affiliation{%
  \institution{University of Science and Technology of China}
  \country{China}
}
\authornote{Corresponding author.}


\begin{abstract}
Click-through rate (CTR) prediction is one of the core tasks in recommender systems. User behavior sequences, as one of the most effective features, can accurately reflect user preferences and significantly improve prediction accuracy. Richer behavior sequences often enable more comprehensive user profiling, and recent studies have shown that scaling the length of user behavior sequence can yield substantial gains in CTR. However, due to the widespread sparsity in recommender systems, incomplete behavior sequences are common in real-world scenarios. Existing sequential modeling methods often rely solely on the target user’s own behavior, and therefore struggle in such scenarios.

This paper proposes a novel method called SUIN (\textbf{S}imilar \textbf{U}sers-augmented \textbf{I}nterest \textbf{N}etwork), which enhances the target user’s behavior sequence with behaviors from similar users to enhance the user profile for CTR prediction. Specifically, we use behavior embeddings encoded by a sequence encoder to retrieve users with similar behaviors from a user retrieval pool. The behavior sequences of these similar users are then concatenated with that of the target user in descending order of similarity to construct an augmented sequence. Given that the augmented sequence contains behaviors from multiple users, we propose a user-specific target-aware position encoding, which identifies the source user of each behavior and captures its relative position to the target item. Furthermore, to mitigate the empirically observed noise in similar users’ behaviors, we design a user-aware target attention that jointly considers item-item and user-user correlations, fully exploiting the potential of the augmented behavior sequence. Comprehensive experiments on widely-used short-term and long-term sequence benchmark datasets demonstrate that our method significantly outperforms state-of-the-art sequential CTR models. 

\end{abstract}


\received{20 February 2007}
\received[revised]{12 March 2009}
\received[accepted]{5 June 2009}

\maketitle

\section{INTRODUCTION}
\begin{figure}[tbp]
  \centering
  \captionsetup[sub]{labelfont=normalfont, textfont=normalfont, skip=-1pt}
  
  \begin{subfigure}[tbp]{\linewidth}
    \centering
    \includegraphics[width=0.9\linewidth]{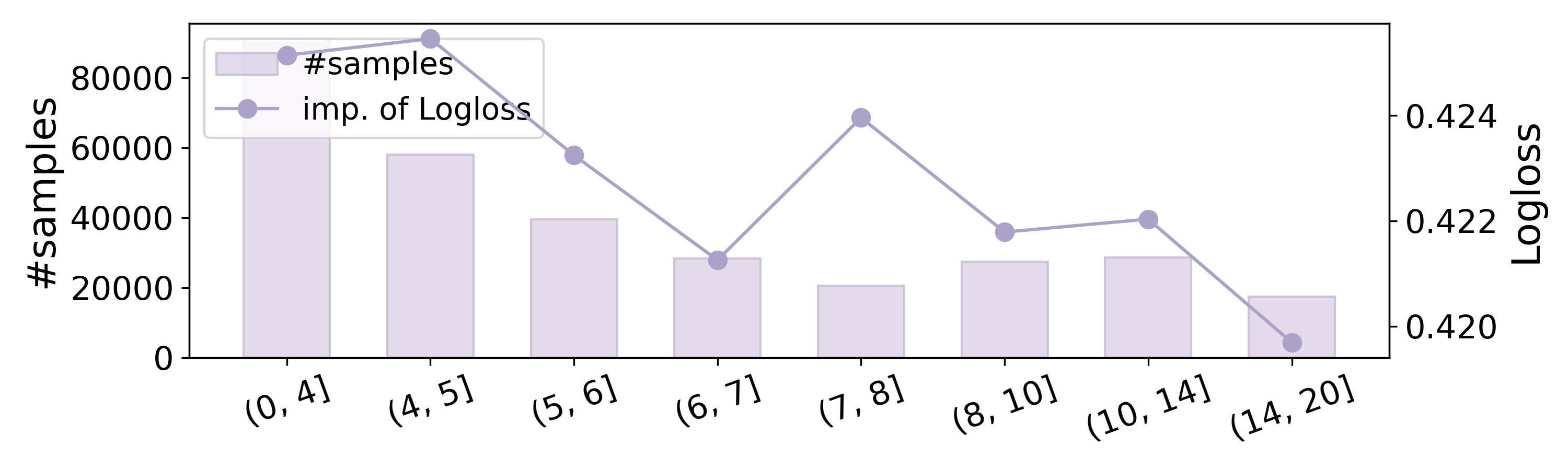}
    \caption{Logloss of DIN on Amazon Electronics}
    \label{fig:motivating_A}
  \end{subfigure}
  
  \vspace{0.2em}
  
  \begin{subfigure}[tbp]{\linewidth}
    \centering
    \includegraphics[width=0.9\linewidth]{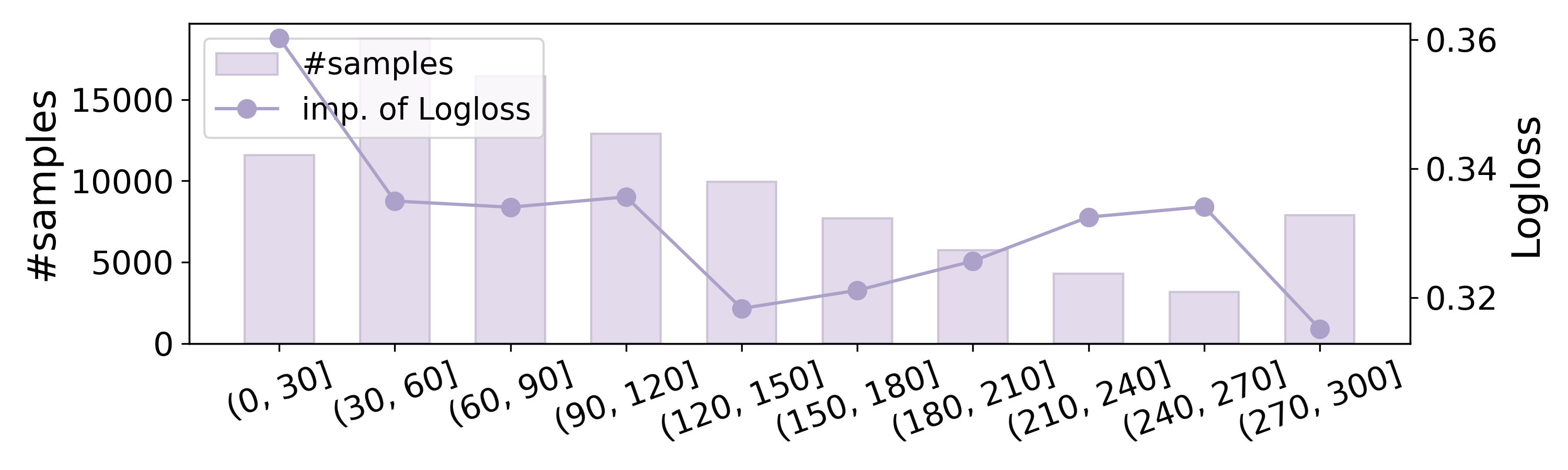}
    \caption{Logloss of TWIN on Taobao}
    \label{fig:motivating_B}
  \end{subfigure}
  \vspace{-5pt}
  \caption{Logloss for user groups with various lengths of behavior sequence on the Amazon Electronics and Taobao. The bar represents the number of users in each group, and the line chart shows the average logloss in each group.}
  \label{fig:motivating}
\vspace{-10pt}
\end{figure}

\begin{figure*}[tbp]
    \centering
    \includegraphics[width=0.9\textwidth]{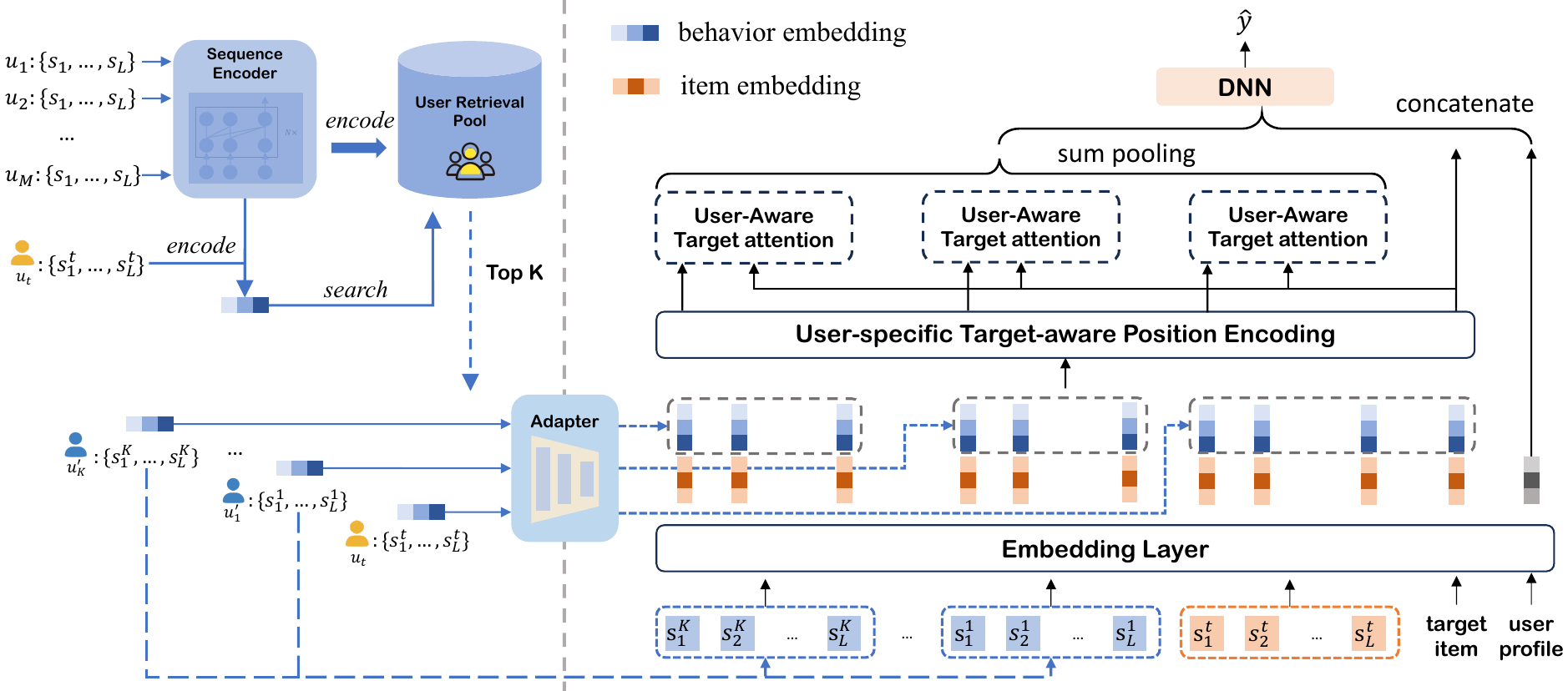}
    \vspace{-5pt}
    \caption{Overall framework of SUIN. The left part depicts similar-user retrieval. Similar users’ behavior sequences are concatenated with the target user’s sequence to form an augmented sequence, which is then modeled by User-specific Target-aware Positional Encoding and User-Aware Target Attention.}
    \label{fig:overall_frame}
\vspace{-8pt}
\end{figure*}
\subsection{Overall Framework}

The rapid explosion of information on e-commerce platforms, content platforms, and advertising services has made recommendation systems an indispensable part of the online ecosystem. A powerful recommendation system (RS) not only enables users to efficiently discover content of interest from vast information pools, but also helps content creators reach potential target audiences, thereby significantly enhancing both user experience and platform revenue.  Click-Through Rate (CTR) prediction, as a fundamental task in recommendation systems, aims to estimate the probability that a user will click on a given item. Its practical significance has made it a central focus of research in both academia and industry over the past few decades, resulting in the emergence of numerous CTR prediction models\cite{rendle2010factorization,juan2016field,chen2016deep,cheng2016wide,guo2017deepfm,wang2017deep,xiao2017attentional,lian2018xdeepfm}.

Existing studies have convincingly demonstrated that a user's historical behavior can effectively capture their interest preferences\cite{wang2019sequential, boka2024survey, pi2019practice}, serving as a ``silver bullet'' in user modeling for CTR prediction. Driven by the rapid advances in deep learning, a variety of user behavior modeling approaches based on different architectures—such as Recurrent Neural Networks (RNNs), self-attention mechanisms, and target attention—have been successively proposed\cite{zhou2018deep, zhou2019deep, feng2019deep, chen2019behavior, zhou2024temporal, xiao2020deep}. 
In recent years, an increasing number of studies have shown that scaling the length of user behavior sequences enables more comprehensive interest modeling and improves CTR prediction accuracy. Two-stage approaches \cite{pi2019practice, pi2020search, chen2021end, cao2022sampling, chang2023twin, si2024twin} leverage users’ lifelong behaviors within practical inference latency by first performing coarse behavior retrieval and then applying fine-grained target-attention modeling. More recent work adopts a request-level amortization strategy \cite{guo2025request} to achieve end-to-end modeling of sequences up to length 10k \cite{guan2025make}. These advances have delivered substantial gains in real-world production systems.
To further validate this phenomenon, we first conduct empirical studies on two public datasets, with the results shown in Figure \ref{fig:motivating}. Specifically, we group users based on the length of their behavior sequences and compute the Logloss metric for each user group, where both short- and long-sequence modeling methods are evaluated (DIN\cite{zhou2018deep} and TWIN\cite{chang2023twin}). The experimental results indicate an overall trend of improved prediction performance as the length of user behavior sequences increases. 
Unfortunately, in real-world scenarios, user behavior histories are not always as extensive as the viewing history on video platforms. In many real-world scenarios, particularly in long-form content platforms such as movies and books, user interactions tend to be sparse. Additionally, user behaviors that occur infrequently but carry high-value feedback--such as purchases or likes--are also prevalent and pose significant modeling challenges \cite{wu2021self,xie2022contrastive,ma2018entire,wei2022contrastive}. Publicly available datasets similarly reveal a long-tail distribution in user behavior sequence lengths, with the majority of users having relatively short behavior histories\cite{huang2024recall}. 
However, existing sequential modeling methods typically rely solely on individual users' own behavior sequences for prediction, often struggling to make accurate estimates when faced with sparse scenarios or users with limited behavior histories.

To effectively enrich user behavior histories for more comprehensive user profiling, we draw inspiration from the Retrieval-Augmented Generation (RAG) paradigm \cite{gao2023retrieval,fan2024survey} that that has become prevalent in the LLM era, and propose a method called \textbf{SUIN} (\textbf{S}imilar \textbf{U}sers-augmented \textbf{I}nterest \textbf{N}etwork), which retrieves similar users for a given target user and incorporates their behavior sequences to augment the target user’s own behavior sequence. The resulting augmented sequence provides a richer context for the CTR model, enabling it to make more accurate predictions. The overall framework is illustrated in Figure~\ref{fig:overall_frame}.
In the context of CTR prediction, the behavior sequence of a given user is treated as a query, while the behavior sequences of all users are regarded as candidates and collectively form the retrieval pool. Specifically, we first employ a sequence encoder to encode all users' behavior sequences and construct the retrieval pool. For each target user, we encode their behavior sequence using the same encoder to obtain the query embedding, and retrieve the top-k most similar users from the pool. The retrieved behavior sequences from similar users are then concatenated with the target user's behavior sequence to form an augmented sequence. Directly utilizing all retrieved sequences without differentiation may introduce noise, potentially degrading model performance. To mitigate this, we integrate the embeddings of both the target user and the retrieved users into the target attention mechanism, introducing a user-aware target attention that dynamically adjusts the contribution weights between the original and retrieved behavior sequences.
Our contributions are summarized as follows:
\begin{itemize}
    \item We propose a novel cross-user context augmentation strategy that leverages behavior sequences from similar users to provide a more comprehensive context for CTR prediction.
    \item We introduce \textbf{SUIN} (\textbf{S}imilar \textbf{U}sers-augmented \textbf{I}nterest \linebreak \textbf{N}etwork), with a user-specific target-aware positional encoding to model positional relations among multi-user behaviors and a user-aware target attention to jointly capture item–item and user–user correlations, thereby fully exploiting the augmented sequence.
    \item We conduct extensive experiments on public short-term and long-term sequence datasets, demonstrating the effectiveness and superiority of the proposed method.
\end{itemize}

\section{PRELIMINARY}

\subsection{Task Formulation}

The CTR prediction task aims to estimate the probability of a user clicking on a given item under a specific context. 
Formally, the objective of the CTR prediction task is to learn a binary classification model $f:\mathbb{R}^d \rightarrow \mathbb{R}$ based on a given training dataset $\mathcal{D} = \{ (\mathbf{x}_i, y_i) |i=1,...,N \}$: 
\begin{equation}
    \hat{y}_i = \sigma(f(\mathbf{x}_i))
\end{equation}
where $\sigma$ denotes the sigmoid function that scales the model output to the range $(0, 1)$, $\mathbf{x}_i \in \mathbb{R}^d$ denotes the feature vector of the $i$-th training sample, typically comprising 
user profile features, behavior history features, contextual features, and target item features, 
and $y_i \in \{0, 1\}$ is the corresponding label ($1$ for click, $0$ for no click).

The model $f$ is commonly trained by minimizing the binary cross-entropy (BCE) loss:
\begin{equation}
    \mathcal{L}_{\mathrm{BCE}} = -\frac{1}{N} \sum_{i=1}^{N} \left[ y_i \log(\hat{y}_i) + (1 - y_i) \log(1 - \hat{y}_i) \right]
\end{equation}

\subsection{Classic CTR Prediction Architecture}
\textit{Embedding\&FeatureInteraction} is a classic paradigm widely adopted by various deep CTR models\cite{zhou2018deep, zhou2019deep, feng2019deep, chen2019behavior, zhou2024temporal}. It typically consists of three components: an embedding layer, a sequence pooling layer, and a feature interaction layer. Our proposed method also follows this established paradigm.

\subsubsection{Embedding layer}
During the data preprocessing stage, numerical features in the input are typically transformed into categorical features. Therefore, we assume that all input features are categorical features. For a categorical feature F, the original high-dimensional sparse one-hot or multi-hot encoding $\mathbf{x}_F \in \{0,1\}^{v_F}$ is usually transformed into a low-dimensional embedding to enable effective processing by deep models:
\begin{equation}
    \mathbf{e}_F = \mathbf{x}_F \mathbf{E}_F
\end{equation}
where $\mathbf{E}_F \in \mathbb{R}^{v_F \times d}$ denotes the embedding table for feature F, $v_F$ is the cardinality of feature F, and $d$ is the embedding dimension.

For convenience of notation, we denote the embedding representation of the user behavior sequence $S = [s_1, s_2, \ldots, s_i, \ldots, s_L]$ as $\mathbf{e}_S = [\mathbf{e}_1, \mathbf{e}_2, \ldots, \mathbf{e}_i, \ldots, \mathbf{e}_L] \in R^{L \times d}$, where $s_i$ represents the feature of the $i$-th behavior in the sequence. The embedding of the target item is denoted as $\mathbf{e}_t$.

\subsubsection{Sequence pooling layer}
To obtain a fixed-length embedding and prevent the sequence embedding from becoming overly high-dimensional, the behavior sequence representation $\mathbf{e}_S \in \mathbb{R}^{L \times d}$ is typically aggregated into a fixed-size vector through a pooling layer, which is then concatenated with other features and fed into the feature interaction layer:
\begin{equation}
    \mathbf{e}_{\text{pooling}} = \text{pooling}(\mathbf{e}_1, \mathbf{e}_2, \ldots, \mathbf{e}_L) \in \mathbb{R}^d
\end{equation}

The sequence pooling layer is a central component of sequence modeling methods. 
The subsequent description of our method will primarily focus on the design of the sequence pooling layer.

\subsubsection{Feature interaction layer}
Taking as input the concatenation of the behavior sequence representation and other feature embeddings, the feature interaction layer is responsible for modeling the interactions among all features. 
Although various designs exist for this component\cite{rendle2010factorization, juan2016field, chen2016deep, cheng2016wide, guo2017deepfm, wang2017deep, lian2018xdeepfm, xiao2017attentional}, our method does not focus on it and simply adopts a multilayer perceptron (MLP) as the feature interaction layer for both our method and all baselines.

\section{METHODOLOGY}
In this section, we provide a detailed introduction to the proposed method. We first present the overall framework, followed by an in-depth description of three key components: construction of the user retrieval pool, behavior sequence augmentation with similar users, and the user-aware target attention.

Our proposed method aims to augment the behavior sequence of a given user by incorporating behavior sequences from similar users, thereby enhancing user profiling and improving CTR prediction performance. To achieve this goal, we design three key components: 
\begin{itemize}[left=0pt]
    \item \textbf{User Retrieval Pool.} Each user's behavior sequence is encoded into a behavior embedding via a sequence encoder, and all users' embeddings collectively form the retrieval pool for efficient retrieval of similar users.
    \item \textbf{Behavior Sequence Augmentation.} For a given user, we retrieve the top-k most similar users from the user retrieval pool and leverage their behavior sequences to augment the behavior sequence of the given user.
    \item \textbf{User-Aware Target Attention.} To adaptively mitigate the noise potentially introduced by similar users, the user-aware target attention is designed to dynamically adjust the attention weights between the original and retrieved behavior sequences with the aid of their behavior embeddings.

\end{itemize}

Through these three components, we effectively enhance the user behavior sequence with similar users and fully exploit the potential of the augmented sequence for CTR prediction. The overall architecture of our proposed method is illustrated in Figure \ref{fig:overall_frame}.

\subsection{User Retrieval Pool}
\subsubsection{Sequence Encoder}
We posit that users whose behavioral patterns resemble that of the target user offer more valuable information for enhancing the target user's profile. Accordingly, we represent each user by their behavior sequence. To enable efficient retrieval of similar users, it is necessary to employ a sequence encoder to transform the user's behavior sequence into a dense behavior embedding.

In the widely adopted multi-stage recommendation architecture, the retrieval stage typically precedes the ranking stage (where CTR models are commonly applied) and serves to narrow down the candidate set\cite{qin2022rankflow,huang2023cooperative,liu2024recflow}. Sequence-based dual-tower models\cite{wang2019sequential,kang2018self,sun2019bert4rec,zhou2022filter}, in which the user tower encodes a user's behavior sequence into a user embedding and the item tower encodes item features into item embeddings, are extensively used in the retrieval stage, with the user tower naturally serving as a powerful sequence encoder. Therefore, we adopt the user tower from typical retrieval sequence-based retrieval models as our sequence encoder, which can be formalized as follows:
\begin{equation}
    \mathbf{e}_b = \mathrm{SE}([s_1, s_2, \dots, s_i, \dots, s_L]) \in \mathbb{R}^{d'}
\end{equation}
where $\mathrm{SE}$ denotes the sequence encoder, which takes the user behavior sequence $[s_1, s_2, \dots, s_i, \dots, s_L]$ as input and outputs a $d'$-dimensional behavior embedding $\mathbf{e}_b$. In this work, we adopt SASRec \cite{kang2018self}, a self-attention–based sequence model, as the sequence encoder, since it is a representative and stable model in sequential recommendation. It is pre-trained using the binary cross-entropy (BCE) loss on training data for subsequent use.

In our framework, the sequence encoder is treated as a replaceable component. Section \ref{sec:diff_encoders} analyzes the performance of SUIN when equipped with different sequence encoders, demonstrating its compatibility. Furthermore, the encoder can, in principle, be replaced by more complex models to incorporate richer item features or scenario-specific factors \cite{xie2022decoupled,lei2023practical,elsayed2024multi,hu2025alphafuse}.

\subsubsection{Construction of User Retrieval Pool}
Using the pre-trained sequence encoder, we encode the behavior sequences of all users to obtain their behavior embeddings, which collectively form the user retrieval pool for subsequent similar-user retrieval:
\begin{equation}
    \mathcal{P} = \{ \mathbf{e}_b^1, \dots, \mathbf{e}_b^i, \dots, \mathbf{e}_b^M \}
\end{equation}
where $\mathcal{P}$ denotes the user retrieval pool, $\mathbf{e}_b^i$ denotes the behavior embedding of user $u_i$ and $M$ is the total number of users.

It is worth noting that, to prevent data leakage, the retrieval pool is constructed from users in the training set throughout all experiments, excluding those in the validation and test sets.

\subsection{Behavior Sequence Augmentation}

\subsubsection{Behavior Sequence Augmentation via Similar Users}
For a given user $u_t$, we utilize a pre-trained sequence encoder to obtain its behavior embedding $e_b^t$. Cosine similarity serves as the relevance score between the target and candidate users’ behavior embeddings, a common practice in recommender systems and dense retrieval \cite{wu2021self,reimers2019sentence,zhang2025qwen3}. Taking a user $u_c$ from the user retrieval pool as an example, let $\mathbf{e}_b^c$ denotes its behavior embedding. The relevance score is calculated as:
\begin{equation}
    \mathrm{Similarity}(u_t, u_c) = \frac{\mathbf{e}_b^t {\mathbf{e}_b^c}^\top}{|\mathbf{e}_b^t| \cdot |\mathbf{e}_b^c|}
\end{equation}
We then select the top-K most similar users from the user retrieval pool, denoted as $\{u_1', u_2', \ldots, u_K'\}$. The retrieval can be computed offline to avoid incurring additional latency during CTR inference.

Inspired by the na\"ive RAG\cite{gao2023retrieval}, which synthesizes retrieved document chunks with the posed query to form an extended prompt, we concatenate the behavior sequences of the top-$K$ similar users with that of the target user, ordered by descending similarity, to construct an augmented behavior sequence:
\begin{equation}
    S_t^{\prime} = \{ s_1^K, s_2^K, \dots, s_L^K, \dots, s_1^1, s_2^1, \dots, s_L^1, s_1^t, s_2^t, \dots, s_L^t \}
\end{equation}

where $\{ s_1^k, s_2^k, \dots, s_L^k \}$ denotes the behavior sequence of the $k$-th most similar user $u_k'$, and $\{ s_1^t, s_2^t, \dots, s_L^t \}$ denotes the behavior sequence of the target user $u_t$.

Besides cosine similarity on dense representations, we further analyze alternative similarity measures in Section \ref{sec:diff_similarity}.

\subsubsection{User-Specific Target-Aware Position Encoding}
\label{sec:UTPE}

\begin{figure}[tbp]
    \centering
    \includegraphics[width=0.9\linewidth]{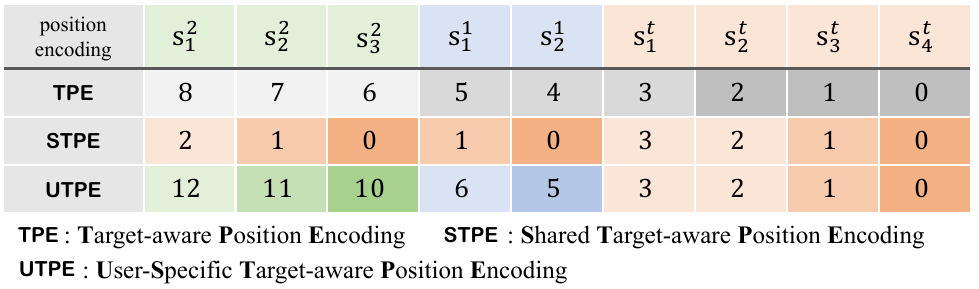}
    \vspace{-5pt}
    \caption{A Toy example illustrating different positional encoding methods. The maximum sequence length is 5, the sequence of target user is $\{s_1^t, s_2^t, s_3^t, s_4^t\}$ and the top-2 similar users are $\{s_1^1, s_2^1\}$ and $\{s_1^2, s_2^2, s_3^2\}$. Colors indicate user identity and lighter shades indicate longer temporal distance. Refer to Section \ref{sec:diff_pos} for details of TPE and STPE.}
    \label{fig:pos_enc}
\vspace{-12pt}
\end{figure}

Temporal information plays a vital role in behavior sequence modeling. Recent studies have highlighted that target-aware position encoding—i.e. position encoding relative to the target item—is an effective technique for improving the performance of sequence modeling\cite{zhou2024temporal,yuan2025contextual}.

To account for the fact that the augmented behavior sequence includes actions from multiple users, we further propose the user-specific target-aware position encoding. For clarity, consider a behavior sequence $S = \{s_1, s_2, \ldots, s_L\}$, where $s_L$ denotes the most recent action. If the sequence length exceeds L, it is truncated; otherwise, it is left-padded.
For the augmented sequence $S_t^{\prime}$, we assign the following position IDs:
\begin{equation}
    \mathrm{POS}_t^{\prime} = \{\dots, \underline{kL + L - 1, \ldots, kL + 1, kL}, \ldots, \underline{L - 1, \ldots, 1, 0} \}
\end{equation}

Specifically, the position IDs assigned to the k-th most similar user's sequence are always $\{ kL + L - 1, \ldots, kL + 1, kL \}$ ($k=0$ is for the target user), which enables the model to identify the source user of each behavior and ensures that more similar users are always assigned smaller position IDs. Moreover, the position ID assigned to the $i$-th latest behavior is always $kL+i- 1$, which is equivalent to relative position encoding to the target item within the $k$-th similar user. Therefore, UTPE (\textbf{U}ser-specific \textbf{T}arget-aware \textbf{P}osition \textbf{E}ncoding) possesses three key properties: (1) awareness of the user each behavior belongs to, (2) awareness of the relative position of behaviors across users, and (3) awareness of the relative position of behaviors within a user.

It is worth noting that UTPE is fundamentally different from applying a target-aware position encoding after concatenating all non-padding behaviors from similar and target users. In the latter case, due to varying sequence lengths, the position IDs of behaviors from the $k$-th similar user become inconsistent, losing the user-specific property. Figure \ref{fig:pos_enc} illustrates a toy example. Detailed explanations of various positional encodings and their comparison with UTPE can be found in Section \ref{sec:diff_pos}.

\subsection{User-Aware Target Attention}

\subsubsection{Challenges in Leveraging Augmented Behavior Sequences}

\begin{table}[tbp]
\caption{AUC results of naive utilization of augmented behavior sequences on DIN and TIN.}
\vspace{-5pt}
\label{tab:naive_aug_seq}
\resizebox{0.75\columnwidth}{!}{%
    \begin{tabular}{cccccc}
    \toprule
    TopK & 0 & 1 & 2 & 3 & 4 \\
    \midrule
    DIN & 0.8833 & 0.8837 & \textbf{0.8854} & 0.8849 & 0.8848 \\
    TIN & \textbf{0.8856} & 0.8846 & 0.8849 & 0.8791 & 0.8785 \\
    \bottomrule
    \end{tabular}%
}
\vspace{-5pt}
\end{table}

\begin{figure}[tbp]
    \centering
    \includegraphics[width=0.85\linewidth]{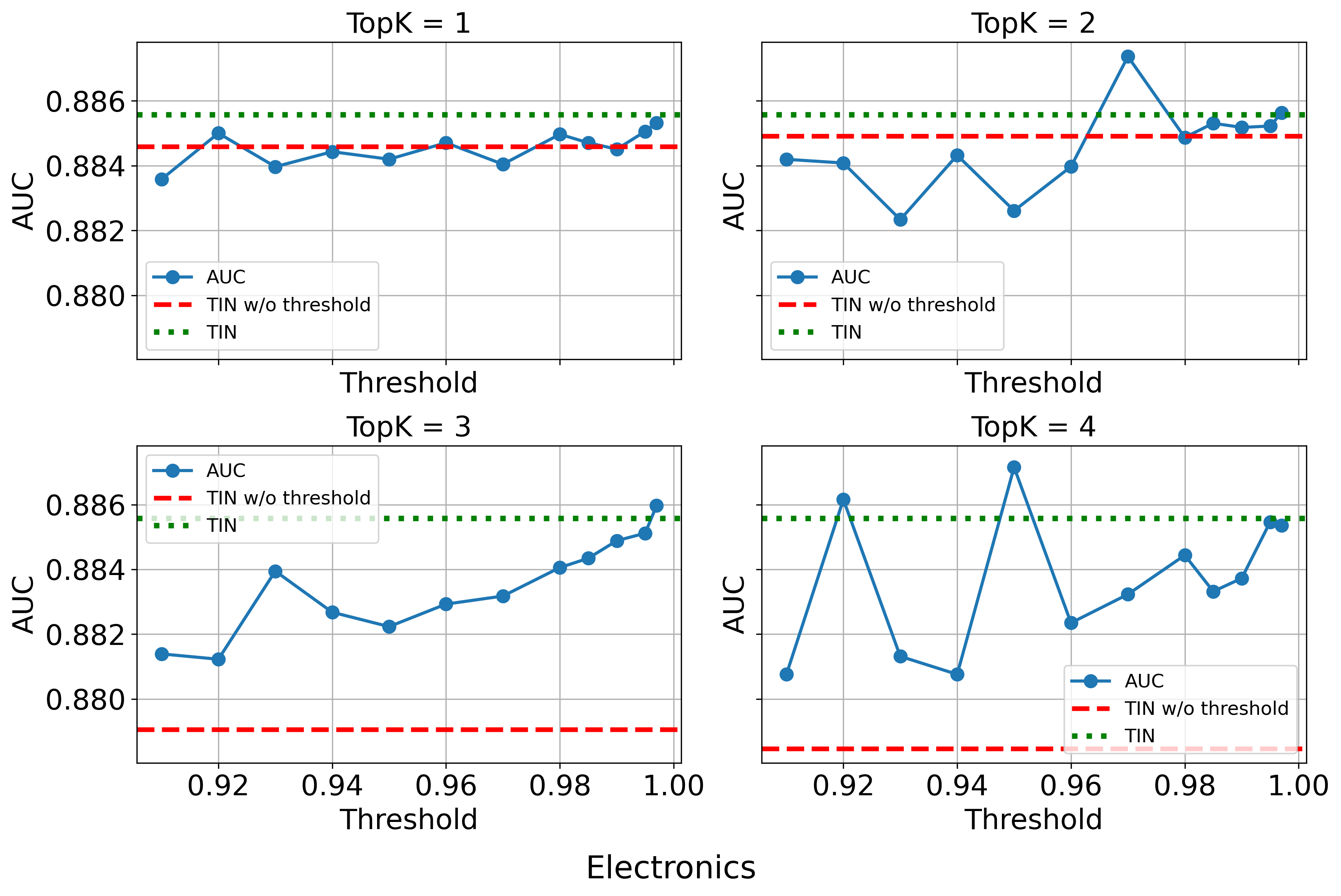}
    \vspace{-5pt}
    \caption{Utilizing augmented behavior sequences with threshold on Electronics.}
    \label{fig:seq_w_threshold}
\vspace{-12pt}
\end{figure}

A straightforward and intuitive approach is to feed the augmented behavior sequence directly into the existing models \cite{zhou2018deep,zhou2024temporal}. However, experimental results indicate that this strategy yields limited effectiveness. As shown in Table \ref{tab:naive_aug_seq}, compared to the backbone model, the performance improvements are marginal, and the TIN model \cite{zhou2024temporal} even experiences a performance drop. Moreover, after reaching a performance peak, further increasing the number of similar users $k$ leads to a decline in performance.

We believe that this is likely due to the substantial noise introduced by the behavior sequences of similar users. To verify this, we filter out users whose similarity scores fall below a predefined threshold. As shown in Figure \ref{fig:seq_w_threshold}, experimental results on the TIN model demonstrate that, under an appropriate threshold, the augmented sequence outperforms the backbone model, providing empirical support for the existence of noise. However, threshold-based filtering still suffers from several limitations. On one hand, it lacks flexibility, discarding entire behavior sequences in a coarse-grained manner, which may restrict performance gains. On the other hand, the threshold value is difficult to tune and often relies on manual heuristics. Therefore, there is a pressing need for a solution that can incorporate the similarity between the target user and retrieved users while enabling fine-grained control over individual behaviors in the augmented sequence.

\subsubsection{Design of User-Aware Target Attention}
\begin{figure}[tbp]
    \centering
    \includegraphics[width=0.75\linewidth]{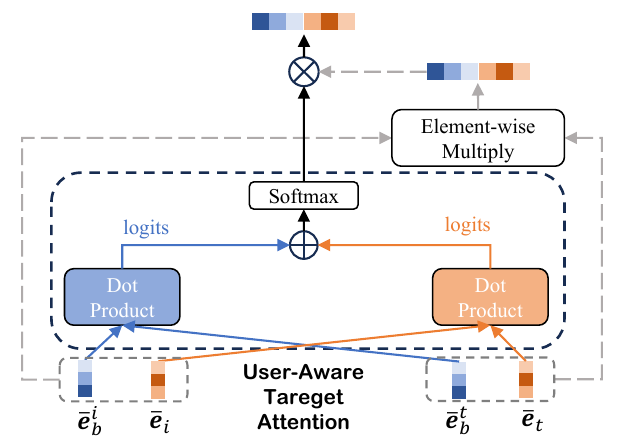}
    \vspace{-5pt}
    \caption{Illustration of the user-aware target attention}
    \label{fig:user_aware_TA}
\vspace{-10pt}
\end{figure}

Compared to standard behavior sequences, augmented behavior sequences incorporate behaviors from multiple users. Thus, the attention weight assigned to each behavior should be determined by two complementary factors: the item-item correlation between the target item and the item of the behavior—which is the primary focus of existing target attention mechanisms—and the user-user correlation between the target user and the similar user associated with the behavior, which is rarely considered in prior work.

To comprehensively assess the importance of behaviors in the augmented sequence, we propose the \textbf{user-aware target attention}. This design incorporates the behavior embedding—generated by the sequence encoder—as an additional input to the target attention, enabling the modeling of user-user correlation, and fuses both item-item and user-user correlations to fine-grainedly control the attention distribution over the augmented sequence, thereby fully exploiting the potential of augmented behavior sequences. The illustration of the user-aware target attention is shown in Figure \ref{fig:user_aware_TA}.

Specifically, a \textbf{behavior embedding adapter} is first employed to project the behavior embeddings of the target user and its similar users into a lower-dimensional space and align them with the parameter space of the CTR prediction model:
\begin{equation}
\tilde{\mathbf{e}}_b = MLP(\mathbf{e}_b)
\end{equation}
where $\mathbf{e}_b$ denotes the behavior embedding of either the target user or a similar user, $\tilde{\mathbf{e}}_b$ is the corresponding projected behavior embedding and $MLP$ is a two-layer Multi-Layer Perceptron with an input dimension of $d'$ and an output dimension of $d$, using ReLU\cite{glorot2011deep} as the activation function. Note that the behavior embeddings are frozen to preserve the sequential information. As a result, they can be precomputed before CTR model training, eliminating the need for online encoding during training.

Similar to TIN\cite{zhou2024temporal}, our target attention module is also built upon scaled dot-product attention. We first describe the modeling of item-item correlation. Let $s_i$ denotes the $i$-th behavior in the augmented sequence $S_t^{\prime}$ with item embedding $\mathbf{e}_i \in \mathbb{R}^{d_{item}}$. The function $p(\cdot)$ returns the position ID assigned by UTPE, and $\mathbf{P}_{item}$ denotes the position embedding layer for item-item correlation. The position-aware item embedding is computed by summing the item and position embeddings: $ \overline{\mathbf{e}}_i = \mathbf{e}_i + \mathbf{P}_{item}^{p(s_i)} $. Similarly, for the target item, we have $ \overline{\mathbf{e}}_t = \mathbf{e}_t + \mathbf{P}_{item}^0 $. The item-item attention logit between behavior $s_i$ and the target item is computed as:
\begin{equation}
    \alpha_{item}^i = \frac{ ({\mathbf{W}_{item}^Q\overline{\mathbf{e}}_t})^\top  (\mathbf{W}_{item}^K\overline{\mathbf{e}}_i) }{\sqrt{d_{item}}}
\end{equation}
where $\mathbf{W}_{item}^Q$ and $\mathbf{W}_{item}^K$ denote the query and key projection matrices for item-item correlation, respectively. 

For user-user correlation, we can also obtain the position-aware behavior embeddings of the user $u_i$ (to which behavior $s_i$ belongs) and the target user $u_t$ as follows:
$\overline{\mathbf{e}}_b^i = \tilde{\mathbf{e}}_b^i + \mathbf{P}_{user}^{p(s_i)}$,
$\overline{\mathbf{e}}_b^t = \tilde{\mathbf{e}}_b^t + \mathbf{P}_{user}^{0}$,
where $\tilde{\mathbf{e}}_b^i$ and $\tilde{\mathbf{e}}_b^t$ are the projected behavior embeddings of $u_i$ and $u_t$, respectively, and $\mathbf{P}_{user}$ denotes the position embedding layer for user-user correlation. The user-user attention logit between $s_i$ and the target user $u_t$ is computed as:
\begin{equation}
    \alpha_{user}^i = \frac{ ({\mathbf{W}_{user}^Q \overline{\mathbf{e}}_b^t})^\top (\mathbf{W}_{user}^K \overline{\mathbf{e}}_b^i) }{ \sqrt{d} }
\end{equation}
where $\mathbf{W}_{user}^Q$ and $\mathbf{W}_{user}^K$ denote the query and key projection matrices for user-user correlation, respectively. 

We integrate the two types of correlation by summing the item-item and user-user attention logits. The final user-aware target attention weight can be formalized as:
\begin{equation}
    \boldsymbol{\alpha} = Softmax(\boldsymbol{\alpha}_{item} + \boldsymbol{\alpha}_{user})
\end{equation}
where
\begin{align*}
\boldsymbol{\alpha}_{item} = [\alpha_{item}^1, \alpha_{item}^2, \ldots, \alpha_{item}^{KL}], \\
\boldsymbol{\alpha}_{user} = [\alpha_{user}^1, \alpha_{user}^2, \ldots, \alpha_{user}^{KL}].
\end{align*}

Using the attention weights $\boldsymbol{\alpha}$, we obtain the aggregated representation of the augmented behavior sequence as follows:
\begin{equation}
   \mathbf{e}_{SUIN} = \alpha_i \cdot \left( [\mathbf{W}_{user}^Q \overline{\mathbf{e}}_t ; \mathbf{W}_{item}^Q \overline{\mathbf{e}}_b^t] \odot [\mathbf{W}_{user}^V \overline{\mathbf{e}}_i ; \mathbf{W}_{item}^V \overline{\mathbf{e}}_b^i] \right) 
\end{equation}
where $\mathbf{W}_{item}^V$ and $\mathbf{W}_{user}^V$ denote the value projection matrices for item-item and user-user correlations respectively, $\odot$ denotes the element-wise product, and $[\cdot ; \cdot]$ represents vector concatenation. Target-aware representation has proven to be an effective trick in TIN \cite{zhou2024temporal}. Following TIN, instead of aggregating $[\mathbf{W}_{user}^V \overline{\mathbf{e}}_i ; \mathbf{W}_{item}^V \overline{\mathbf{e}}_b^i]$, we also aggregate $\left( [\mathbf{W}_{user}^Q \overline{\mathbf{e}}_t ; \mathbf{W}_{item}^Q \overline{\mathbf{e}}_b^t] \odot [\mathbf{W}_{user}^V \overline{\mathbf{e}}_i ; \mathbf{W}_{item}^V \overline{\mathbf{e}}_b^i] \right)$ to better leverage the interaction between the target and behaviors.

\section{EXPERIMENTS}
This section begins by introducing the experimental setup. Section \ref{sec:overall_performance} compares the overall performance of SUIN against various baselines, while Section \ref{sec:ablation} conducts ablation studies to analyze contributions of its three key modules. Section \ref{sec:hyper_param} investigates the influence of the number of similar users. Finally, Section \ref{sec:further_analysis} further examines model compatibility and compares module variants, followed by a group analysis.

\subsection{Experimental Setup}
\subsubsection{Datasets}
\begin{table}[tbp]
  \centering
  \caption{Statistics of the datasets after preprocessing.}
  \vspace{-7pt}
  \label{tab:datasets}
  \resizebox{0.9\columnwidth}{!}{
      \begin{tabular}{cccccc}
        \toprule
         & \textbf{Datasets} & \textbf{\#Users} & \textbf{\#Items} & \textbf{\#Inters} & \textbf{Avg Len} \\
        \midrule
        \multirow{3}{*}{Short} & Electronics & 1,641,026 & 368,228 & 15,473,536 & 9 \\
         & Kindle Store & 892,164 & 466,576 & 16,070,783 & 18 \\
        \midrule
        \multirow{2}{*}{Long} & Taobao & 987,994 & 4,162,024 & 100,150,807 & 101 \\
         & Alipay & 498,308 & 2,200,291 & 35,179,371 & 70 \\
        \bottomrule
      \end{tabular}%
    }
\vspace{-10pt}
\end{table}

To assess the effectiveness of the proposed method, we conduct experiments on publicly available short-term and long-term sequence datasets. Table \ref{tab:datasets} summarizes the statistics of the processed datasets.

\begin{itemize}[leftmargin=*]
    \item \textbf{Short-term sequence datasets}: \textbf{Amazon\footnote{\url{https://amazon-reviews-2023.github.io}}} consists of multiple subsets containing large-scale product reviews and metadata, and is widely used for short-term sequence modeling. We select two subsets—\textbf{Electronics} and \textbf{Kindle Store}—for evaluation. Following the setting in \cite{zhou2019deep}, the prediction task is defined as determining whether a user will leave a review for a given item.
    The datasets are split 8:1:1 by users for training, validation, and testing.
    For each user, the most recent review is treated as a positive sample, and one item different from it is randomly sampled as a negative sample.
    
    \item \textbf{Long-term sequence datasets}: 
    Long-term sequence datasets typically feature extended user behavior sequences for evaluating long-term modeling.
    Commonly used long-term datasets sequence include \textbf{Taobao}\footnote{\url{https://tianchi.aliyun.com/dataset/649}} and \textbf{Alipay}\footnote{\url{https://tianchi.aliyun.com/dataset/dataDetail?dataId=53}}. \textbf{Taobao} contains real interaction logs collected from the Taobao mobile app during Nov 25 to Dec 3, 2017. \textbf{Alipay} covers online payment behaviors from Jul 1 to Nov 30, 2015. For these datasets, we split the data into training, validation, and test sets in an 8:1:1 ratio. Negative samples are generated via random sampling, with a 1:1 ratio of positive to negative samples. A maximum sequence length of 300 is used for evaluation.
\end{itemize}

\subsubsection{Baselines}
To validate the effectiveness of our proposed method, we compare it with a range of classical and state-of-the-art models under both short-term and long-term sequence CTR prediction settings. The short-term CTR prediction baselines include \textbf{Avg-Pooling}, \textbf{DIN}\cite{zhou2018deep}, \textbf{BST}\cite{chen2019behavior}, \textbf{DIEN}\cite{zhou2019deep}, \textbf{DSIN} \cite{feng2019deep}, \textbf{DMIN}\cite{xiao2020deep}, and \textbf{TIN}\cite{zhou2024temporal}. The long-term CTR prediction baselines include \textbf{SIM-hard}\cite{pi2020search}, \textbf{SIM-soft}\cite{pi2020search}, \textbf{ETA}\cite{chen2021end}, \textbf{SDIM}\cite{cao2022sampling}, and \textbf{TWIN}\cite{chang2023twin}.

\subsubsection{Evaluation Metrics}
We adopt two widely used metrics in the CTR prediction field—AUC (Area Under the Curve) and Logloss (cross-entropy loss)—as our evaluation criteria.

\subsubsection{Implementation Details}
For all datasets, we adopt SASRec as the sequence encoder and train it using the binary cross-entropy (BCE) loss with a random negative sample. To ensure fairness and prevent data leakage, the sequence encoder is trained with the same data split and the same amount of data as the CTR prediction model.
Across all datasets and CTR models, the embedding dimension is set to 16, and the feature interaction layers use a DNN with hidden sizes [200, 80, 1], with ReLU as the activation function. For training, we adopt the Adam optimizer with a learning rate of 0.001, apply early stopping with a patience of 1, and set the maximum number of epochs to 5. The batch size is set to 256 for long-term sequence datasets and 512 for short-term sequence datasets. All models are implemented based on FuxiCTR\cite{zhu2021open}, and the hyperparameters of baseline models follow the configurations recommended in the original papers. For SUIN, the behavior embedding adapter uses hidden sizes of [32, 16] with ReLU activations, and its dropout rate is tuned over [0, 0.1, 0.2, 0.5]. The optimal Top-K on the four datasets is 4, 2, 2, and 1, respectively. In the long-term sequence setting, SUIN adopts a strategy similar to TWIN by sharing parameters between CP-GSU and ESU to enable two-stage modeling.

\subsection{Overall Performance}
\label{sec:overall_performance}

\begin{table}[tbp]
\caption{Results on short-term sequence datasets. The best and the second-best performances are denoted in bold and underlined fonts, respectively.}
\vspace{-5pt}
\label{tab:short_term_auc}
\resizebox{0.9 \columnwidth}{!}{%
    \begin{tabular}{ccccc}
        \toprule
        \multirow{2}{*}{model} & \multicolumn{2}{c}{Electronics} & \multicolumn{2}{c}{Kindle Store} \\  
        \cmidrule(lr){2-3} \cmidrule(lr){4-5}  & AUC $\uparrow$ & logloss $\downarrow$ & AUC $\uparrow$ & logloss $\downarrow$ \\
        \midrule
        SASRec & 0.8750 & 0.4691 & 0.8750 & 1.3271 \\
        \midrule
        Avg-Pooling & 0.8793 & 0.4319 & 0.8974 & 0.4048 \\
        DIN & 0.8833 & 0.4287 & 0.8910 & 0.4150 \\
        BST & 0.8862 & 0.4256 & 0.8952 & 0.4238 \\
        DIEN & \underline{0.8873} & \underline{0.4187} & 0.8977 & 0.4059 \\
        DSIN & 0.8849 & 0.4275 & 0.8900 & 0.4197 \\
        DMIN & 0.8859 & 0.4273 & 0.8866 & 0.4299 \\
        TIN & 0.8856 & 0.4282 & \underline{0.9002} & \underline{0.3988} \\
        \midrule
        SUIN & \textbf{0.8911} & \textbf{0.4132} & \textbf{0.9068} & \textbf{0.3857} \\ 			
        $\Delta\%$ & +0.42\% & +1.31\% & +0.73\% & +3.29\% \\
        \bottomrule
    \end{tabular}%
}
\vspace{-8pt}
\end{table}

\begin{table}[tbp]
\caption{Results on long-term sequence datasets. The best and the second-best performances are denoted in bold and underlined fonts, respectively.}
\vspace{-5pt}
\label{tab:long_term_auc}
\resizebox{0.9\columnwidth}{!}{%
    \begin{tabular}{ccccc}
        \toprule
        \multirow{2}{*}{model} & \multicolumn{2}{c}{Taobao} & \multicolumn{2}{c}{Alipay} \\
        \cmidrule(lr){2-3} \cmidrule(lr){4-5}  & AUC $\uparrow$ & logloss $\downarrow$ & AUC $\uparrow$ & logloss $\downarrow$ \\
        \midrule
        SASRec & 0.8056 & 1.2106 & 0.8170 & 0.6749 \\
        \midrule
        Avg-Pooling & 0.8807 & 0.4327 & 0.8384 & 0.4890 \\
        SIM-hard & 0.9252 & 0.3476 & 0.8718 & 0.4461 \\
        SIM-soft & \underline{0.9339} & \underline{0.3259} & 0.9031 & 0.3885 \\
        ETA & 0.9091 & 0.3819 & 0.853 & 0.4719 \\
        SDIM & 0.9070 & 0.3848 & 0.8775 & 0.4377 \\
        TWIN & 0.9314 & 0.3328 & \underline{0.9056} & \underline{0.3818} \\
        \midrule
        SUIN & \textbf{0.9384} & \textbf{0.3165} & \textbf{0.9121} & \textbf{0.3669} \\
        $\Delta\%$ & +0.48\% & +2.90\% & +0.72\% & +3.90\% \\
        \bottomrule
    \end{tabular}%
}
\vspace{-10pt}
\end{table}

Tables \ref{tab:short_term_auc} and \ref{tab:long_term_auc} report the experimental results on the short- and long-term sequence datasets, respectively. Across all datasets, SASRec underperforms compared to other CTR models due to its dual-tower architecture, which fails to adequately capture the interaction between the target item and the user context. This observation also helps rule out the possibility that the improvements of our method stem solely from a powerful sequence encoder.

On short-term datasets, a series of improvements over DIN—such as BST, DIEN, DSIN, DMIN, and TIN—achieve varying degrees of performance gains compared to DIN. Among them, DIEN stands out as the strongest baseline on the Electronics dataset, while TIN serves as the best-performing baseline on the Kindle Store dataset.

On long-term datasets, SIM-soft leverages an embedding-based dot-product relevance score between the target and behaviors, while TWIN employs a powerful and consistent target attention across two stages. These two methods achieve the second-best performance on the Taobao and Alipay datasets, respectively.

Finally, benefiting from the similar-user-augmented behavior sequences and the user-aware target attention, SUIN consistently outperforms all baselines across both short- and long-term settings. Specifically, in terms of AUC, SUIN achieves relative improvements of +0.42\% on Electronics, +0.73\% on Kindle Store, +0.48\% on Taobao, and +0.72\% on Alipay compared to the best-performing baselines. These results clearly demonstrate the effectiveness of our proposed method.

\subsection{Ablation Study}
\label{sec:ablation}

\begin{figure}[tbp]
    \centering
    \includegraphics[width=0.85\linewidth]{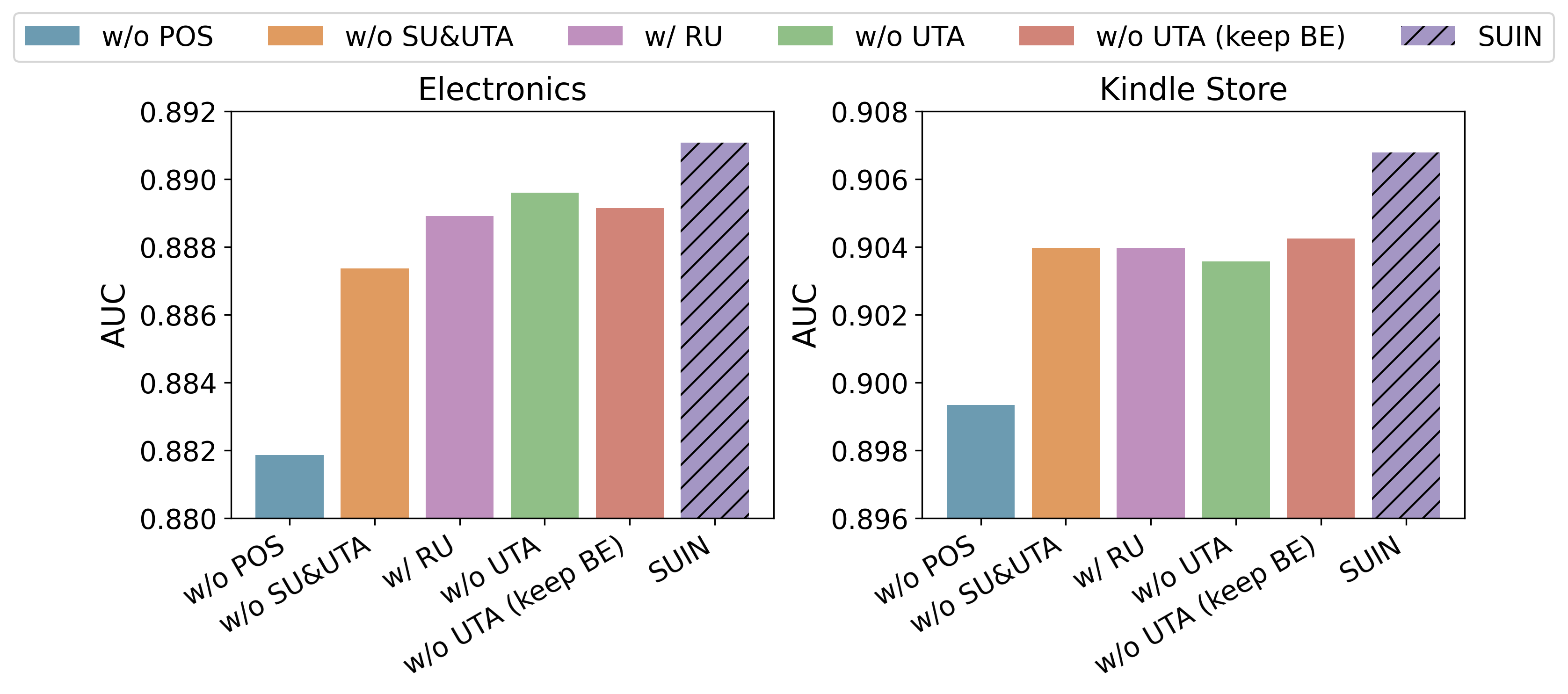}
    \vspace{-5pt}
    \caption{Ablation study on Electronics and Kindle Store. "POS" denotes position embedding, "SU" denotes similar users, "RU" denotes random sampled users, "UTA" denotes user-aware target attention, and "BE" denotes the behavior embeddings produced by the sequence encoder.}
    \label{fig:ablation}
\vspace{-15pt}
\end{figure}

The similar users-augmented behavior sequences, the UTPE-based position encoding for enhanced behavior sequences, and the user-aware target attention constitute the three key components of our proposed method. To investigate their effectiveness, we design the following variants:

\begin{itemize}[leftmargin=*]
\item \textbf{w/o UTA}: Removes the user-aware target attention.
\item \textbf{w/o UTA (keep BE)}: Removes the user-aware target attention, while still using the projected behavior embeddings of the target user and similar users from the sequence encoder as additional features for the feature interaction layer.
\item \textbf{w/ RU}: Replaces similar users with randomly sampled users while keeping all other modules unchanged.
\item \textbf{w/o SU\&UTA}: Removes both similar-user augmentation and the user-aware target attention, degenerating into a standard target-attention model.
\item \textbf{w/o POS}: Removes the UTPE position embedding used for augmented behavior sequences.
\end{itemize}

The experimental results are shown in Figure \ref{fig:ablation}. \textbf{w/o UTA} exhibits a noticeable performance drop, indicating that user-aware target attention plays a key role in effectively leveraging the similar-user-augmented behavior sequences. The performance of \textbf{w/o UTA (keep BE)} remains close to \textbf{w/o UTA}, suggesting that the advantage of UTA lies not in simply incorporating behavior embeddings, but in its architectural design that integrates both item-item and user-user correlations.

For similar users–augmented behavior sequences, both \textbf{w/ RU} and \textbf{w/o SU\&UTA} lead to substantial performance degradation, indicating that the contextual information provided by similar users is a key driver of the performance gains. Furthermore, the deterioration under w/ RU suggests that the improvement is not merely due to architecture modifications, but rather to the synergy between the architecture and similar-user augmentation.

\textbf{w/o POS} achieves the worst results. Without position embeddings, the model cannot identify the relative positions between behaviors from the target and similar users, nor distinguish the user each behavior belongs to, making it highly vulnerable to the noise introduced by the similar-user behavior sequences.

\subsection{Hyper-parameter Analysis}
\label{sec:hyper_param}

\begin{figure}[tbp]
    \centering
    \includegraphics[width=0.85\linewidth]{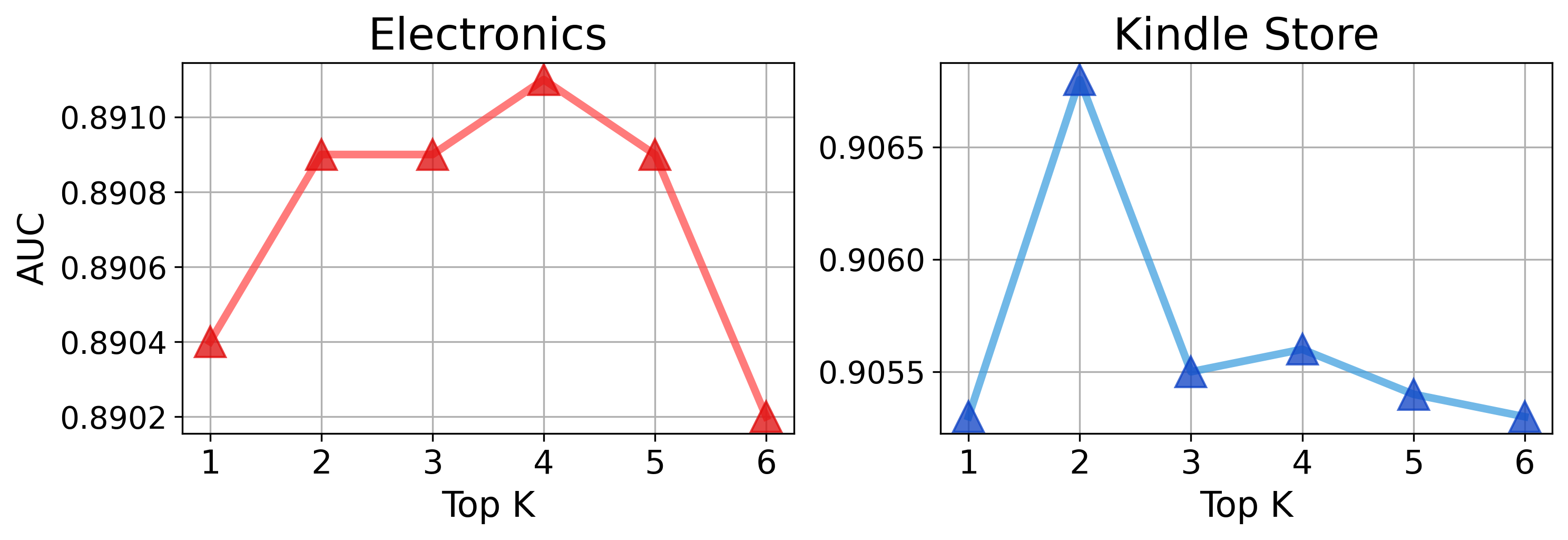}
    \vspace{-5pt}
    \caption{Performance of SUIN with 1 to 6 similar users on Amazon Electronics and Kindle Store.}
    \label{fig:sim_user_topk}
\vspace{-8pt}
\end{figure}

The number of similar users is a critical hyperparameter of SUIN. We investigate its influence on model performance and offer guidance for hyperparameter selection in broader application scenarios. Figure \ref{fig:sim_user_topk} illustrates the variation of AUC when SUIN is configured with different numbers of similar users on the Amazon Electronics and Kindle Store datasets. The results exhibit a pronounced unimodal trend on both datasets as the number of similar users increases. Specifically, the optimal performance is attained with 4 similar users on the Electronics dataset, while 2 similar users lead to the best performance on the Kindle Store dataset. This observation is well-motivated: in the regime of few similar users, expanding the set of similar users provides richer context for CTR prediction and improves performance, while an excessive number of similar users tends to introduce noise and leads to performance degradation. Notably, even after surpassing the optimal setting and experiencing a performance drop,  SUIN still outperforms the baseline without similar-user augmentation.

\subsection{Further Analysis}
\label{sec:further_analysis}

\subsubsection{Compatibility with other sequence encoders}
\label{sec:diff_encoders}

\begin{table}[tbp]
\caption{Performance of SUIN equipped with different behavior sequence.}
\vspace{-5pt}
\resizebox{0.85\columnwidth}{!}{%
\begin{tabular}{ccccc}
    \toprule
     & \multicolumn{2}{c}{Electronics} & \multicolumn{2}{c}{Kindle Store} \\
    \cmidrule(lr){2-3} \cmidrule(lr){4-5} \multirow{-2}{*}{Encoder} & AUC $\uparrow$ & logloss $\downarrow$ & AUC $\uparrow$ & logloss $\downarrow$ \\
    \midrule
    - & 0.8874 & 0.4201 & 0.9040 & 0.3931 \\
    \midrule
    GRU4Rec & 0.8907 & 0.4132 & 0.9045 & 0.3925 \\
    SASRec & \textbf{0.8911} & 0.4132 & \textbf{0.9068} & \textbf{0.3857} \\
    BERT4Rec & 0.8910 & \textbf{0.4124} & 0.9057 & 0.3869 \\
    \bottomrule
\end{tabular}%
}
\label{tab:diff_encoders}
\vspace{-8pt}
\end{table}

We further analyze the performance of SUIN equipped with different sequence encoders. The experimental results in Table \ref{tab:diff_encoders} show that SUIN with all encoders consistently outperforms the backbone without similar-user enhancement (i.e., \textit{SUIN w/o SU\&UTA} in Section \ref{sec:ablation}), demonstrating the compatibility of our framework. SASRec \cite{kang2018self} and BERT4Rec \cite{sun2019bert4rec} are both attention-based models and achieve comparable performance, and both surpass the RNN-based GRU4Rec \cite{hidasi2015session}. This observation aligns with the evolution of sequential recommendation models and indicates that more expressive sequence encoders can better capture user behavioral patterns, thereby yielding further gains in CTR prediction.

\subsubsection{Compatibility with other user similarity measures}
\label{sec:diff_similarity}

\begin{table}[tbp]
\caption{Performance of different similarity measures}
\vspace{-5pt}
\resizebox{0.85\columnwidth}{!}{%
\begin{tabular}{ccccc}
    \toprule
    \multirow{2}{*}{Measure} & \multicolumn{2}{c}{Electronics} & \multicolumn{2}{c}{Kindle Store} \\
    \cmidrule(lr){2-3} \cmidrule(lr){4-5} & AUC $\uparrow$ & logloss $\downarrow$ & AUC $\uparrow$ & logloss $\downarrow$ \\
    \midrule
    - & 0.8874 & 0.4201 & 0.9040 & 0.3931 \\
    \midrule
    Cosine & \textbf{0.8911} & \textbf{0.4132} & 0.9068 & 0.3857 \\
    Inner Product & 0.8910 & 0.4136 & 0.9059 & 0.3874 \\
    Euclidean & 0.8909 & 0.4132 & 0.9052 & 0.3897 \\
    \midrule
    Jaccard & 0.8901 & 0.4150 & 0.9065 & 0.3865 \\
    User-Swing & 0.8901 & 0.4144 & \textbf{0.9071} & \textbf{0.3851} \\
    \bottomrule
\end{tabular}%
}
\label{tab:diff_similarity}
\vspace{-8pt}
\end{table}

We evaluate the compatibility of SUIN with different user similarity measures, as shown in Table \ref{tab:diff_similarity}. Dense measures include cosine similarity, inner product and Euclidean distance, while the co-occurrence–based statistical methods include Jaccard similarity \cite{stitini2022investigating}, as well as User-Swing, an adaptation of Swing \cite{yang2020large} for user similarity.

Using any similarity measure yields performance improvements over the backbone (i.e., \textit{SUIN w/o SU\&UTA} in Section \ref{sec:ablation}), indicating that SUIN is compatible with diverse similarity metrics. Dense measures exhibit similar performance, with cosine similarity being the most stable across datasets. Statistical methods perform slightly worse than cosine similarity on Electronics but perform comparably on Kindle Store, with User-Swing marginally outperforming cosine similarity. We attribute this to the fact that statistical methods rely on an additional sequence encoder to obtain behavior embeddings and can thus be interpreted as an ensemble of statistical and deep representation methods. Overall, cosine similarity demonstrates the most stable performance across datasets and achieves the best trade-off between effectiveness and cost.

\subsubsection{Comparison with other position encoding methods}
\label{sec:diff_pos}

\begin{table}[tbp]
\caption{Performance of different position encoding}
\vspace{-5pt}
\resizebox{0.85\columnwidth}{!}{%
\begin{tabular}{cccccc}
    \toprule
     \multirow{2}{*}{\makecell{position \\ encoding}} & \multirow{2}{*}{properties} & \multicolumn{2}{c}{Electronics} & \multicolumn{2}{c}{Kindle Store} \\
    \cmidrule(lr){3-4} \cmidrule(lr){5-6} & & AUC $\uparrow$ & logloss $\downarrow$ & AUC $\uparrow$ & logloss $\downarrow$ \\
    \midrule
    UTPE & \textcolor{green!50!black}{\ding{51}} \textcolor{green!50!black}{\ding{51}} \textcolor{green!50!black}{\ding{51}} & \textbf{0.8911} & \textbf{0.4132} & \textbf{0.9068} & \textbf{0.3857} \\
    TPE & \textcolor{red!60!black}{\ding{55}} \textcolor{green!50!black}{\ding{51}} \textcolor{green!50!black}{\ding{51}} & 0.8893 & 0.4162 & 0.9051 & 0.3879 \\
    STPE & \textcolor{red!60!black}{\ding{55}} \textcolor{red!60!black}{\ding{55}} \textcolor{green!50!black}{\ding{51}} & 0.8840 & 0.4244 & 0.9006 & 0.3996 \\
    None & \textcolor{red!60!black}{\ding{55}} \textcolor{red!60!black}{\ding{55}} \textcolor{red!60!black}{\ding{55}} & 0.8819 & 0.4276 & 0.8993 & 0.3990 \\
    \bottomrule
\end{tabular}%
}
\label{tab:diff_pos}
\vspace{-12pt}
\end{table}

Table \ref{tab:diff_pos} compares \textbf{UTPE} (Section \ref{sec:UTPE}) with several positional encoding variants. We identify three desired properties for position encoding methods: (1) awareness of the user each behavior belongs to, (2) awareness of the relative position of behaviors across users, and (3) awareness of the relative position of behaviors within a user. In the `properties' column of Table 5, a check mark indicates whether a position encoding method possesses the corresponding property. \textbf{UTPE} satisfies all three.

\textbf{TPE} refers to a naive target-aware position encoding, where non-padding behaviors from both the target user and similar users are concatenated before applying target-aware encoding. Since lengths of user sequences vary, the encoding of the $k$-th most similar user is not fixed, and thus it lacks awareness of the user each behavior belongs to. \textbf{STPE} denotes a shared target-aware position encoding in which all users share a single set of target-aware position embeddings, retaining only awareness of the relative position of behaviors within a user. \textbf{None} represents the setting without any position embedding. 
Experimental results show that \textbf{UTPE} consistently outperforms other position encoding methods.

\subsubsection{Performance improvement across different sequence augmentation ratios}

\begin{figure}[tbp]
    \centering
    \includegraphics[width=0.8\linewidth]{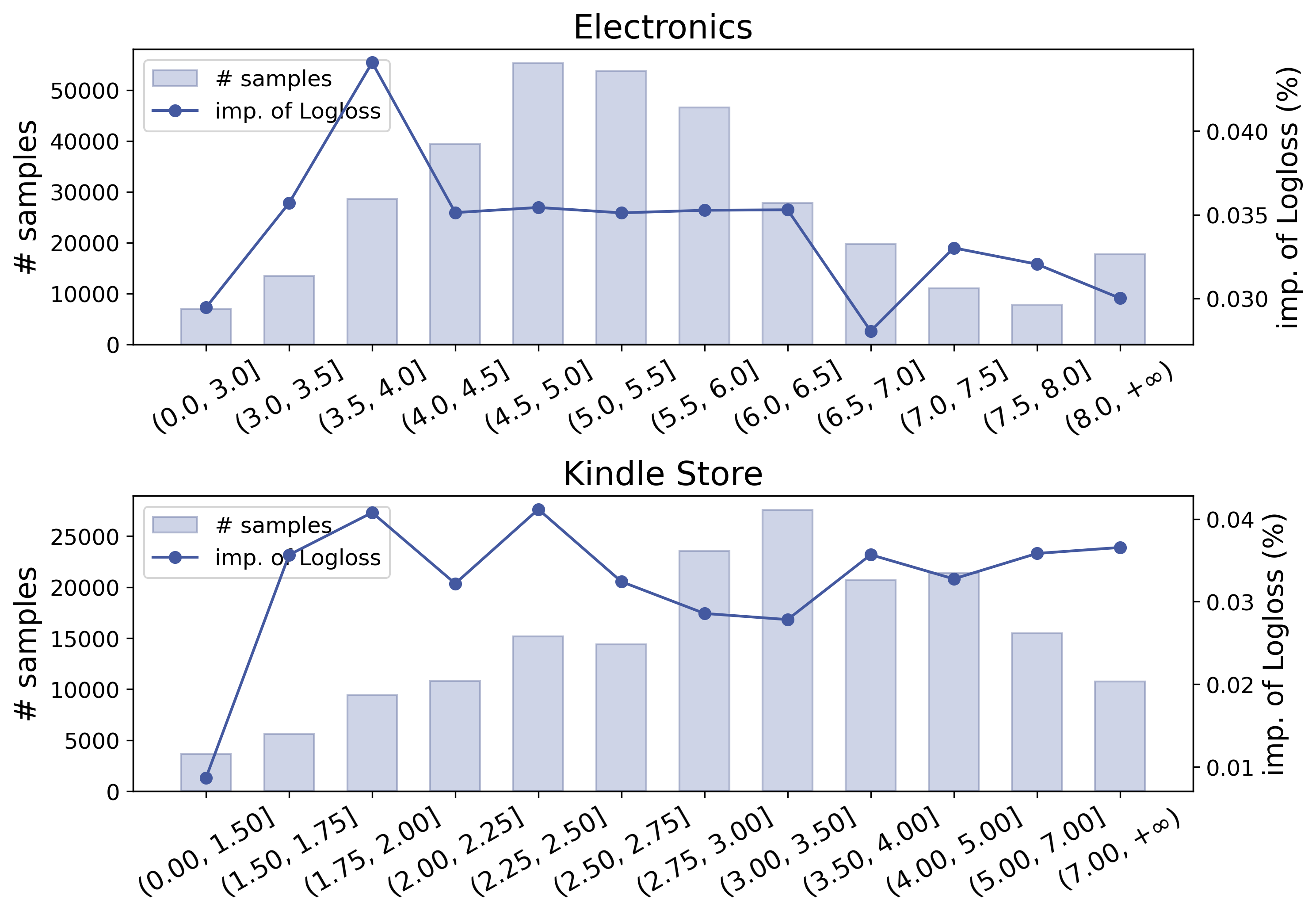}
    \vspace{-5pt}
    \caption{Performance improvement over TIN across different sequence augmentation ratios (x-axis) on Electronics and Kindle Store.}
    \label{fig:imp_wrt_merge_rate}
\vspace{-10pt}
\end{figure}

We define the sequence augmentation ratio as the ratio of the augmented sequence length to the original sequence length. To investigate the effectiveness of the proposed method under varying augmentation ratios, we group users by augmentation ratio and report group-wise Logloss. The results on the Electronics and Kindle Store datasets are shown in Figure \ref{fig:imp_wrt_merge_rate}. When the augmentation ratio is low, the performance improvement of our method over the baseline gradually increases as the ratio grows, indicating that the augmented behavior sequences contribute positively to the model's effectiveness. After reaching a peak, further increasing the augmentation ratio leads to a decline or fluctuation in improvement. We attribute this to the growing amount of noise introduced by excessive augmentation.  Nevertheless, our method still yields gains across all user groups.

\section{RELATED WORK}
\subsection{Click-Through Rate Prediction}

Click-through rate (CTR) prediction is a core task in recommendation systems. Feature interaction constitutes a key component in this task, and research has evolved from early factorization machine–based models \cite{rendle2010factorization,juan2016field} to advanced architectures capable of capturing high-order interactions \cite{chen2016deep,guo2017deepfm,wang2017deep,xiao2017attentional,lian2018xdeepfm}.

In recent years, 
user behavior sequence modeling has attracted increasing attention. Models such as DIN \cite{zhou2018deep}, DIEN \cite{zhou2019deep}, DSIN \cite{feng2019deep} and DMIN \cite{xiao2020deep} demonstrated the effectiveness of target attention (TA) and interest modeling. To better exploit long sequences, MIMN \cite{pi2019practice} introduced the user interest center architecture based on an offline-updated memory matrix. Subsequent studies adopted a two-stage cascade framework, in which a General Search Unit (GSU) retrieves the top-k behaviors most relevant to the candidate item from the long-term sequence, followed by an Exact Search Unit (ESU) performing fine-grained TA modeling. Among them, SIM and UBR4CTR retrieve behaviors within the same category, while ETA and SDIM apply locality-sensitive hashing and multi-round hashing as the GSU, respectively. TWIN \cite{chang2023twin} and TWIN-v2 \cite{si2024twin} further introduce a consistency-preserved GSU to mitigate the inconsistency between the two stages. These studies collectively demonstrate that rich behavior sequences and effective sequence modeling play a crucial role in CTR prediction.

Several existing CTR prediction studies adopt similar-user augmentation strategies related to ours, primarily by constructing additional user–user edges on user–item interaction graphs. Specifically, NI-CTR \cite{min2022neighbour} connects similar users via multi-hop paths in a heterogeneous graph, while DG-ENN \cite{guo2021dual} derives user-user relations based on user preferences and attributes. However, these approaches treat users merely as nodes in a graph and fail to explicitly exploit the fine-grained preference information embedded in similar users’ behavior sequences, leaving their sequence-level augmentation potential unexplored.

\subsection{Retrieval-Augmented Learning}

Retrieval-augmented learning has become an effective paradigm in natural language processing (NLP) \cite{patrick2020retrieval,borgeaud2022improving,izacard2023atlas,gao2023retrieval,fan2024survey} and computer vision (CV) \cite{chen2022re,blattmann2022retrieval}. The core idea is to retrieve relevant information from external databases to enrich the current context, thereby enhancing the model’s performance.

The RS community has also begun exploring the potential of retrieval-augmented learning for CTR prediction. RIM \cite{qin2021retrieval} is a pioneering work that retrieves similar samples via discrete retrieval and aggregates them by feature fields. DERT \cite{zheng2023dense} improves the discrete retrieval in RIM with an embedding-based dense retrieval. PET \cite{du2022learning} and RAT \cite{li2024rat} further improve the interaction modeling between retrieved samples and the target sample: PET constructs all samples into a hypergraph and employs a GNN to model sample relations, whereas RAT utilizes a Transformer with cascaded attention to jointly capture both intra-sample and cross-sample patterns. Different from sample-level retrieval, RAR \cite{huang2024recall} retrieves similar users/items based on user/item-ID, and models user/item-ID interactions within the retrieved sets to enhance CTR prediction.

However, whether through sample-level retrieval or user-ID-based similar user retrieval, these methods overlook explicit behavioral sequence features that are crucial for CTR prediction, leading to the absence of preference signals and sequential patterns in retrieved results. In contrast, our method focuses on sequence-level retrieval and introduces a novel model architecture to fully exploit the retrieved behavioral sequences, thereby addressing the limitations of existing methods.

\section{CONCLUSION}
In this paper, we propose a novel similar users-based sequence augmentation strategy to provide richer contextual information for CTR prediction. Specifically, we retrieve similar users based on dense embeddings encoded by a sequence encoder, and concatenate their behaviors with those of the target user to construct an augmented behavior sequence. To fully exploit the potential of the augmented sequences, we further design a user-specific target-aware positional encoding and a user-aware target attention mechanism. 
Extensive experiments on both short-term and long-term public sequence datasets demonstrate the effectiveness of our method. Ablation studies and further analyses corroborate the contributions of each module. 
For future work, we plan to incorporate multimodal item features into similar-user retrieval and investigate how to effectively fuse similarity users across different modalities. Another promising direction is to provide theoretical explanations for the effectiveness of the proposed UTPE and UTA modules.
\clearpage
\bibliographystyle{ACM-Reference-Format}
\bibliography{main}

@article{guo2025request,
  title={Request-Only Optimization for Recommendation Systems},
  author={Guo, Liang and Li, Wei and Liao, Lucy and Cheng, Huihui and Zhang, Rui and Shi, Yu and Wang, Yueming and Huang, Yanzun and Zhai, Keke and Wang, Pengchao and others},
  journal={arXiv preprint arXiv:2508.05640},
  year={2025}
}

@article{guan2025make,
  title={Make It Long, Keep It Fast: End-to-End 10k-Sequence Modeling at Billion Scale on Douyin},
  author={Guan, Lin and Yang, Jia-Qi and Zhao, Zhishan and Zhang, Beichuan and Sun, Bo and Luo, Xuanyuan and Ni, Jinan and Li, Xiaowen and Qi, Yuhang and Fan, Zhifang and others},
  journal={arXiv preprint arXiv:2511.06077},
  year={2025}
}

@inproceedings{xie2022contrastive,
  title={Contrastive learning for sequential recommendation},
  author={Xie, Xu and Sun, Fei and Liu, Zhaoyang and Wu, Shiwen and Gao, Jinyang and Zhang, Jiandong and Ding, Bolin and Cui, Bin},
  booktitle={2022 IEEE 38th international conference on data engineering (ICDE)},
  pages={1259--1273},
  year={2022},
  organization={IEEE}
}

@inproceedings{ma2018entire,
  title={Entire space multi-task model: An effective approach for estimating post-click conversion rate},
  author={Ma, Xiao and Zhao, Liqin and Huang, Guan and Wang, Zhi and Hu, Zelin and Zhu, Xiaoqiang and Gai, Kun},
  booktitle={The 41st International ACM SIGIR Conference on Research \& Development in Information Retrieval},
  pages={1137--1140},
  year={2018}
}

@inproceedings{wei2022contrastive,
  title={Contrastive meta learning with behavior multiplicity for recommendation},
  author={Wei, Wei and Huang, Chao and Xia, Lianghao and Xu, Yong and Zhao, Jiashu and Yin, Dawei},
  booktitle={Proceedings of the fifteenth ACM international conference on web search and data mining},
  pages={1120--1128},
  year={2022}
}

@inproceedings{glorot2011deep,
  title={Deep sparse rectifier neural networks},
  author={Glorot, Xavier and Bordes, Antoine and Bengio, Yoshua},
  booktitle={Proceedings of the fourteenth international conference on artificial intelligence and statistics},
  pages={315--323},
  year={2011},
  organization={JMLR Workshop and Conference Proceedings}
}

@article{liu2024recflow,
  title={RecFlow: An Industrial Full Flow Recommendation Dataset},
  author={Liu, Qi and Zheng, Kai and Huang, Rui and Li, Wuchao and Cai, Kuo and Chai, Yuan and Niu, Yanan and Hui, Yiqun and Han, Bing and Mou, Na and others},
  journal={arXiv preprint arXiv:2410.20868},
  year={2024}
}

@inproceedings{qin2022rankflow,
  title={Rankflow: Joint optimization of multi-stage cascade ranking systems as flows},
  author={Qin, Jiarui and Zhu, Jiachen and Chen, Bo and Liu, Zhirong and Liu, Weiwen and Tang, Ruiming and Zhang, Rui and Yu, Yong and Zhang, Weinan},
  booktitle={Proceedings of the 45th International ACM SIGIR Conference on Research and Development in Information Retrieval},
  pages={814--824},
  year={2022}
}

@inproceedings{huang2023cooperative,
  title={Cooperative retriever and ranker in deep recommenders},
  author={Huang, Xu and Lian, Defu and Chen, Jin and Zheng, Liu and Xie, Xing and Chen, Enhong},
  booktitle={Proceedings of the ACM Web Conference 2023},
  pages={1150--1161},
  year={2023}
}

@inproceedings{xie2022decoupled,
  title={Decoupled side information fusion for sequential recommendation},
  author={Xie, Yueqi and Zhou, Peilin and Kim, Sunghun},
  booktitle={Proceedings of the 45th international ACM SIGIR conference on research and development in information retrieval},
  pages={1611--1621},
  year={2022}
}

@inproceedings{lei2023practical,
  title={Practical Content-aware Session-based Recommendation: deep retrieve then shallow rank},
  author={Lei, Yuxuan and Chen, Xiaolong and Lian, Defu and Zhang, Peiyan and Lian, Jianxun and Li, Chaozhuo and Xie, Xing},
  booktitle={Amazon KDD Cup 2023 Workshop},
  year={2023}
}

@inproceedings{elsayed2024multi,
  title={Multi-Behavioral Sequential Recommendation},
  author={Elsayed, Shereen and Rashed, Ahmed and Schmidt-Thieme, Lars},
  booktitle={Proceedings of the 18th ACM conference on recommender systems},
  pages={902--906},
  year={2024}
}

@inproceedings{hu2025alphafuse,
  title={Alphafuse: Learn id embeddings for sequential recommendation in null space of language embeddings},
  author={Hu, Guoqing and Zhang, An and Liu, Shuo and Cai, Zhibo and Yang, Xun and Wang, Xiang},
  booktitle={Proceedings of the 48th International ACM SIGIR Conference on Research and Development in Information Retrieval},
  pages={1614--1623},
  year={2025}
}

@article{wang2019sequential,
  title={Sequential recommender systems: challenges, progress and prospects},
  author={Wang, Shoujin and Hu, Liang and Wang, Yan and Cao, Longbing and Sheng, Quan Z and Orgun, Mehmet},
  journal={arXiv preprint arXiv:2001.04830},
  year={2019}
}

@article{hidasi2015session,
  title={Session-based recommendations with recurrent neural networks},
  author={Hidasi, Bal{\'a}zs and Karatzoglou, Alexandros and Baltrunas, Linas and Tikk, Domonkos},
  journal={arXiv preprint arXiv:1511.06939},
  year={2015}
}

@inproceedings{kang2018self,
  title={Self-attentive sequential recommendation},
  author={Kang, Wang-Cheng and McAuley, Julian},
  booktitle={2018 IEEE international conference on data mining (ICDM)},
  pages={197--206},
  year={2018},
  organization={IEEE}
}

@inproceedings{sun2019bert4rec,
  title={BERT4Rec: Sequential recommendation with bidirectional encoder representations from transformer},
  author={Sun, Fei and Liu, Jun and Wu, Jian and Pei, Changhua and Lin, Xiao and Ou, Wenwu and Jiang, Peng},
  booktitle={Proceedings of the 28th ACM international conference on information and knowledge management},
  pages={1441--1450},
  year={2019}
}

@inproceedings{zhou2022filter,
  title={Filter-enhanced MLP is all you need for sequential recommendation},
  author={Zhou, Kun and Yu, Hui and Zhao, Wayne Xin and Wen, Ji-Rong},
  booktitle={Proceedings of the ACM web conference 2022},
  pages={2388--2399},
  year={2022}
}

@article{boka2024survey,
  title={A survey of sequential recommendation systems: Techniques, evaluation, and future directions},
  author={Boka, Tesfaye Fenta and Niu, Zhendong and Neupane, Rama Bastola},
  journal={Information Systems},
  volume={125},
  pages={102427},
  year={2024},
  publisher={Elsevier}
}

@article{reimers2019sentence,
  title={Sentence-bert: Sentence embeddings using siamese bert-networks},
  author={Reimers, Nils and Gurevych, Iryna},
  journal={arXiv preprint arXiv:1908.10084},
  year={2019}
}

@article{zhang2025qwen3,
  title={Qwen3 Embedding: Advancing Text Embedding and Reranking Through Foundation Models},
  author={Zhang, Yanzhao and Li, Mingxin and Long, Dingkun and Zhang, Xin and Lin, Huan and Yang, Baosong and Xie, Pengjun and Yang, An and Liu, Dayiheng and Lin, Junyang and others},
  journal={arXiv preprint arXiv:2506.05176},
  year={2025}
}

@misc{patrick2020retrieval,
    title={Retrieval-Augmented Generation for Knowledge-Intensive NLP Tasks},
    author={Patrick Lewis and Ethan Perez and Aleksandra Piktus and Fabio Petroni and Vladimir Karpukhin and Naman Goyal and Heinrich Küttler and Mike Lewis and Wen-tau Yih and Tim Rocktäschel and Sebastian Riedel and Douwe Kiela},
    year={2020},
    eprint={2005.11401},
    archivePrefix={arXiv},
    primaryClass={cs.CL}
}

@inproceedings{borgeaud2022improving,
  title={Improving language models by retrieving from trillions of tokens},
  author={Borgeaud, Sebastian and Mensch, Arthur and Hoffmann, Jordan and Cai, Trevor and Rutherford, Eliza and Millican, Katie and Van Den Driessche, George Bm and Lespiau, Jean-Baptiste and Damoc, Bogdan and Clark, Aidan and others},
  booktitle={International conference on machine learning},
  pages={2206--2240},
  year={2022},
  organization={PMLR}
}

@article{izacard2023atlas,
  title={Atlas: Few-shot learning with retrieval augmented language models},
  author={Izacard, Gautier and Lewis, Patrick and Lomeli, Maria and Hosseini, Lucas and Petroni, Fabio and Schick, Timo and Dwivedi-Yu, Jane and Joulin, Armand and Riedel, Sebastian and Grave, Edouard},
  journal={Journal of Machine Learning Research},
  volume={24},
  number={251},
  pages={1--43},
  year={2023}
}

@article{gao2023retrieval,
  title={Retrieval-augmented generation for large language models: A survey},
  author={Gao, Yunfan and Xiong, Yun and Gao, Xinyu and Jia, Kangxiang and Pan, Jinliu and Bi, Yuxi and Dai, Yixin and Sun, Jiawei and Wang, Haofen and Wang, Haofen},
  journal={arXiv preprint arXiv:2312.10997},
  volume={2},
  number={1},
  year={2023}
}

@inproceedings{fan2024survey,
  title={A survey on rag meeting llms: Towards retrieval-augmented large language models},
  author={Fan, Wenqi and Ding, Yujuan and Ning, Liangbo and Wang, Shijie and Li, Hengyun and Yin, Dawei and Chua, Tat-Seng and Li, Qing},
  booktitle={Proceedings of the 30th ACM SIGKDD conference on knowledge discovery and data mining},
  pages={6491--6501},
  year={2024}
}

@article{chen2022re,
  title={Re-imagen: Retrieval-augmented text-to-image generator},
  author={Chen, Wenhu and Hu, Hexiang and Saharia, Chitwan and Cohen, William W},
  journal={arXiv preprint arXiv:2209.14491},
  year={2022}
}

@article{blattmann2022retrieval,
  title={Retrieval-augmented diffusion models},
  author={Blattmann, Andreas and Rombach, Robin and Oktay, Kaan and M{\"u}ller, Jonas and Ommer, Bj{\"o}rn},
  journal={Advances in Neural Information Processing Systems},
  volume={35},
  pages={15309--15324},
  year={2022}
}

@article{qin2021retrieval,
  title={Retrieval \& Interaction Machine for Tabular Data Prediction},
  author={Jiarui Qin and Weinan Zhang and Rong Su and Zhirong Liu and Weiwen Liu and Ruiming Tang and Xiuqiang He and Yong Yu},
  journal={Proceedings of the 27th ACM SIGKDD Conference on Knowledge Discovery \& Data Mining},
  year={2021},
}

@inproceedings{zheng2023dense,
author = {Zheng, Lei and Li, Ning and Chen, Xianyu and Gan, Quan and Zhang, Weinan},
title = {Dense Representation Learning and Retrieval for Tabular Data Prediction},
year = {2023},
publisher = {Association for Computing Machinery},
pages = {3559–3569},
series = {KDD '23}
}

@article{du2022learning,
  title={Learning enhanced representation for tabular data via neighborhood propagation},
  author={Du, Kounianhua and Zhang, Weinan and Zhou, Ruiwen and Wang, Yangkun and Zhao, Xilong and Jin, Jiarui and Gan, Quan and Zhang, Zheng and Wipf, David P},
  journal={Advances in neural information processing systems},
  volume={35},
  pages={16373--16384},
  year={2022}
}

@inproceedings{li2024rat,
author = {Li, Yushen and Wang, Jinpeng and Dai, Tao and Zhu, Jieming and Yuan, Jun and Zhang, Rui and Xia, Shu-Tao},
title = {RAT: Retrieval-Augmented Transformer for Click-Through Rate Prediction},
year = {2024},
isbn = {9798400701726},
publisher = {Association for Computing Machinery},
booktitle = {Companion Proceedings of the ACM Web Conference 2024},
pages = {867–870},
series = {WWW '24}
}

@inproceedings{guo2021dual,
  title={Dual graph enhanced embedding neural network for CTR prediction},
  author={Guo, Wei and Su, Rong and Tan, Renhao and Guo, Huifeng and Zhang, Yingxue and Liu, Zhirong and Tang, Ruiming and He, Xiuqiang},
  booktitle={Proceedings of the 27th ACM SIGKDD conference on knowledge discovery \& data mining},
  pages={496--504},
  year={2021}
}

@inproceedings{min2022neighbour,
  title={Neighbour interaction based click-through rate prediction via graph-masked transformer},
  author={Min, Erxue and Rong, Yu and Xu, Tingyang and Bian, Yatao and Luo, Da and Lin, Kangyi and Huang, Junzhou and Ananiadou, Sophia and Zhao, Peilin},
  booktitle={Proceedings of the 45th international ACM SIGIR conference on research and development in information retrieval},
  pages={353--362},
  year={2022}
}

@inproceedings{huang2024recall,
  title={Recall-Augmented Ranking: Enhancing Click-Through Rate Prediction Accuracy with Cross-Stage Data},
  author={Huang, Junjie and Cai, Guohao and Zhu, Jieming and Dong, Zhenhua and Tang, Ruiming and Zhang, Weinan and Yu, Yong},
  booktitle={Companion Proceedings of the ACM Web Conference 2024},
  pages={830--833},
  year={2024}
}

@inproceedings{zhu2021open,
  title={Open benchmarking for click-through rate prediction},
  author={Zhu, Jieming and Liu, Jinyang and Yang, Shuai and Zhang, Qi and He, Xiuqiang},
  booktitle={Proceedings of the 30th ACM international conference on information \& knowledge management},
  pages={2759--2769},
  year={2021}
}

@inproceedings{xiao2020deep,
  title={Deep multi-interest network for click-through rate prediction},
  author={Xiao, Zhibo and Yang, Luwei and Jiang, Wen and Wei, Yi and Hu, Yi and Wang, Hao},
  booktitle={Proceedings of the 29th ACM international conference on information \& knowledge management},
  pages={2265--2268},
  year={2020}
}

@inproceedings{chen2019behavior,
  title={Behavior sequence transformer for e-commerce recommendation in alibaba},
  author={Chen, Qiwei and Zhao, Huan and Li, Wei and Huang, Pipei and Ou, Wenwu},
  booktitle={Proceedings of the 1st international workshop on deep learning practice for high-dimensional sparse data},
  pages={1--4},
  year={2019}
}

@inproceedings{yuan2025contextual,
  title={A Contextual-Aware Position Encoding for Sequential Recommendation},
  author={Yuan, Jun and Cai, Guohao and Dong, Zhenhua},
  booktitle={Companion Proceedings of the ACM on Web Conference 2025},
  pages={577--585},
  year={2025}
}

@inproceedings{zhou2024temporal,
  title={Temporal Interest Network for User Response Prediction},
  author={Zhou, Haolin and Pan, Junwei and Zhou, Xinyi and Chen, Xihua and Jiang, Jie and Gao, Xiaofeng and Chen, Guihai},
  booktitle={Companion Proceedings of the ACM Web Conference 2024},
  pages={413--422},
  year={2024}
}

@inproceedings{zhou2018deep,
  title={Deep interest network for click-through rate prediction},
  author={Zhou, Guorui and Zhu, Xiaoqiang and Song, Chenru and Fan, Ying and Zhu, Han and Ma, Xiao and Yan, Yanghui and Jin, Junqi and Li, Han and Gai, Kun},
  booktitle={Proceedings of the 24th ACM SIGKDD international conference on knowledge discovery \& data mining},
  pages={1059--1068},
  year={2018}
}

@inproceedings{zhou2019deep,
  title={Deep interest evolution network for click-through rate prediction},
  author={Zhou, Guorui and Mou, Na and Fan, Ying and Pi, Qi and Bian, Weijie and Zhou, Chang and Zhu, Xiaoqiang and Gai, Kun},
  booktitle={Proceedings of the AAAI conference on artificial intelligence},
  volume={33},
  number={01},
  pages={5941--5948},
  year={2019}
}

@inproceedings{feng2019deep,
author = {Feng, Yufei and Lv, Fuyu and Shen, Weichen and Wang, Menghan and Sun, Fei and Zhu, Yu and Yang, Keping},
title = {Deep session interest network for click-through rate prediction},
year = {2019},
publisher = {AAAI Press},
booktitle = {Proceedings of the 28th International Joint Conference on Artificial Intelligence},
pages = {2301–2307},
location = {Macao, China},
series = {IJCAI'19}
}

@inproceedings{pi2019practice,
author = {Pi, Qi and Bian, Weijie and Zhou, Guorui and Zhu, Xiaoqiang and Gai, Kun},
title = {Practice on Long Sequential User Behavior Modeling for Click-Through Rate Prediction},
year = {2019},
isbn = {9781450362016},
publisher = {Association for Computing Machinery},
booktitle = {Proceedings of the 25th ACM SIGKDD International Conference on Knowledge Discovery \& Data Mining},
pages = {2671–2679},
series = {KDD '19}
}

@inproceedings{pi2020search,
  title={Search-based user interest modeling with lifelong sequential behavior data for click-through rate prediction},
  author={Pi, Qi and Zhou, Guorui and Zhang, Yujing and Wang, Zhe and Ren, Lejian and Fan, Ying and Zhu, Xiaoqiang and Gai, Kun},
  booktitle={Proceedings of the 29th ACM International Conference on Information \& Knowledge Management},
  pages={2685--2692},
  year={2020}
}

@article{chen2021end,
  title={End-to-end user behavior retrieval in click-through rateprediction model},
  author={Chen, Qiwei and Pei, Changhua and Lv, Shanshan and Li, Chao and Ge, Junfeng and Ou, Wenwu},
  journal={arXiv preprint arXiv:2108.04468},
  year={2021}
}

@inproceedings{cao2022sampling,
  title={Sampling is all you need on modeling long-term user behaviors for CTR prediction},
  author={Cao, Yue and Zhou, Xiaojiang and Feng, Jiaqi and Huang, Peihao and Xiao, Yao and Chen, Dayao and Chen, Sheng},
  booktitle={Proceedings of the 31st ACM International Conference on Information \& Knowledge Management},
  pages={2974--2983},
  year={2022}
}

@inproceedings{chang2023twin,
  title={TWIN: TWo-stage interest network for lifelong user behavior modeling in CTR prediction at kuaishou},
  author={Chang, Jianxin and Zhang, Chenbin and Fu, Zhiyi and Zang, Xiaoxue and Guan, Lin and Lu, Jing and Hui, Yiqun and Leng, Dewei and Niu, Yanan and Song, Yang and others},
  booktitle={Proceedings of the 29th ACM SIGKDD Conference on Knowledge Discovery and Data Mining},
  pages={3785--3794},
  year={2023}
}

@inproceedings{si2024twin,
title = {TWIN V2: Scaling Ultra-Long User Behavior Sequence Modeling for Enhanced CTR Prediction at Kuaishou},
author = {Si, Zihua and Guan, Lin and Sun, Zhongxiang and Zang, Xiaoxue and Lu, Jing and Hui, Yiqun and Cao, Xingchao and Yang, Zeyu and Zheng, Yichen and Leng, Dewei and Zheng, Kai and Zhang, Chenbin and Niu, Yanan and Song, Yang and Gai, Kun},
year = {2024},
publisher = {Association for Computing Machinery},
pages = {4890–4897},
series = {CIKM '24}
}

@inproceedings{rendle2010factorization,
author={Rendle, Steffen},
booktitle={2010 IEEE International Conference on Data Mining}, 
title={Factorization Machines}, 
year={2010},
pages={995-1000}
}

@inproceedings{juan2016field,
author = {Juan, Yuchin and Zhuang, Yong and Chin, Wei-Sheng and Lin, Chih-Jen},
title = {Field-aware Factorization Machines for CTR Prediction},
year = {2016},
isbn = {9781450340359},
publisher = {Association for Computing Machinery},
booktitle = {Proceedings of the 10th ACM Conference on Recommender Systems},
pages = {43–50},
series = {RecSys '16}
}

@inproceedings{chen2016deep,
author = {Chen, Junxuan and Sun, Baigui and Li, Hao and Lu, Hongtao and Hua, Xian-Sheng},
title = {Deep CTR Prediction in Display Advertising},
year = {2016},
publisher = {Association for Computing Machinery},
booktitle = {Proceedings of the 24th ACM International Conference on Multimedia},
pages = {811–820},
series = {MM '16}
}

@inproceedings{cheng2016wide,
author = {Cheng, Heng-Tze and Koc, Levent and Harmsen, Jeremiah and Shaked, Tal and Chandra, Tushar and Aradhye, Hrishi and Anderson, Glen and Corrado, Greg and Chai, Wei and Ispir, Mustafa and Anil, Rohan and Haque, Zakaria and Hong, Lichan and Jain, Vihan and Liu, Xiaobing and Shah, Hemal},
title = {Wide \& Deep Learning for Recommender Systems},
year = {2016},
isbn = {9781450347952},
publisher = {Association for Computing Machinery},
booktitle = {Proceedings of the 1st Workshop on Deep Learning for Recommender Systems},
pages = {7–10},
series = {DLRS 2016}
}

@inproceedings{guo2017deepfm,
author = {Guo, Huifeng and Tang, Ruiming and Ye, Yunming and Li, Zhenguo and He, Xiuqiang},
title = {DeepFM: a factorization-machine based neural network for CTR prediction},
year = {2017},
publisher = {AAAI Press},
booktitle = {Proceedings of the 26th International Joint Conference on Artificial Intelligence},
pages = {1725–1731},
series = {IJCAI'17}
}

@inproceedings{wang2017deep,
author = {Wang, Ruoxi and Fu, Bin and Fu, Gang and Wang, Mingliang},
title = {Deep \& Cross Network for Ad Click Predictions},
year = {2017},
publisher = {Association for Computing Machinery},
booktitle = {Proceedings of the ADKDD'17},
articleno = {12},
series = {ADKDD'17}
}

@inproceedings{lian2018xdeepfm,
author = {Lian, Jianxun and Zhou, Xiaohuan and Zhang, Fuzheng and Chen, Zhongxia and Xie, Xing and Sun, Guangzhong},
title = {xDeepFM: Combining Explicit and Implicit Feature Interactions for Recommender Systems},
year = {2018},
isbn = {9781450355520},
publisher = {Association for Computing Machinery},
booktitle = {Proceedings of the 24th ACM SIGKDD International Conference on Knowledge Discovery \& Data Mining},
pages = {1754–1763},
series = {KDD '18}
}

@inproceedings{xiao2017attentional,
author = {Xiao, Jun and Ye, Hao and He, Xiangnan and Zhang, Hanwang and Wu, Fei and Chua, Tat-Seng},
title = {Attentional factorization machines: learning the weight of feature interactions via attention networks},
year = {2017},
isbn = {9780999241103},
publisher = {AAAI Press},
pages = {3119–3125},
series = {IJCAI'17}
}

@inproceedings{wu2021self,
  title={Self-supervised graph learning for recommendation},
  author={Wu, Jiancan and Wang, Xiang and Feng, Fuli and He, Xiangnan and Chen, Liang and Lian, Jianxun and Xie, Xing},
  booktitle={Proceedings of the 44th international ACM SIGIR conference on research and development in information retrieval},
  pages={726--735},
  year={2021}
}

@article{stitini2022investigating,
  title={Investigating different similarity metrics used in various recommender systems types: Scenario cases},
  author={Stitini, Oumaima and Kaloun, Soulaimane and Bencharef, Omar},
  journal={The International Archives of the Photogrammetry, Remote Sensing and Spatial Information Sciences},
  volume={48},
  pages={187--193},
  year={2022},
  publisher={Copernicus GmbH}
}

@article{yang2020large,
  title={Large scale product graph construction for recommendation in e-commerce},
  author={Yang, Xiaoyong and Zhu, Yadong and Zhang, Yi and Wang, Xiaobo and Yuan, Quan},
  journal={arXiv preprint arXiv:2010.05525},
  year={2020}
}


\end{document}